\documentstyle[12pt,myssi,epsf]{article}

\begin{document}
\thispagestyle{empty}
\setcounter{page}{0}

\begin{flushright}
CERN-TH/97-55\\
March 1997\\
hep-ph/9703429
\end{flushright}
\vspace*{1cm}

\title{Non-relativistic effective theory for 
Quarkonium Production in Hadron Collisions}

\author{Martin Beneke\footnote{
Lecture at the XXIVth SLAC Summer Institute on Particle Physics, 
`The Strong Interaction, from Hadrons to Partons', August 1996.}\\ 
Theory Division, CERN\\
CH -- 1211 Gen\`{e}ve 23, Switzerland\\[0.4cm]
}

\maketitle
\begin{abstract}
I review recent progress in understanding 
inclusive quarkonium production 
in hadron collisions. The first part focuses on 
non-rela\-ti\-vi\-stic QCD as an effective theory. I discuss its 
differences from and similarities with effective theories describing 
bound states of a single heavy quark, as far as matching calculations 
beyond tree-level and power counting are concerned. The second part 
summarizes predictions for charmonium and bottomonium 
production at collider and fixed target experiments and their comparison 
with data. The emphasis here is on novel signatures due to color octet 
production, polarization of quarkonia and the $\chi_1/\chi_2$ ratio 
in fixed target collisions.
\end{abstract}

\setcounter{footnote}{0}

\section{Introduction}

The discovery of charmonium bound states \cite{charmonium} 
opened the world 
of heavy flavours, its wonderful variety and complexity. 
Since then, the focus has shifted to decays 
of bound states of a single heavy quark, their implications for CP 
violation and, perhaps, `New Physics'. Quarkonia, due to their 
leptonic decay signature, on the other hand have become important 
tagging modes in hadron collisions and may become so for the 
quark-gluon plasma phase. For a theorist, therefore, the interest in 
quarkonia is mostly of intrinsic nature. The experimental data are 
there, and we are challenged to explain them.

Quarkonia are the `atoms' of the strong force. If their Rydberg energy 
were larger than $\Lambda$, the dynamical low-energy scale of the 
strong interaction, quarkonia would be weakly coupled bound states 
of a heavy quark and antiquark, tractable in perturbation theory just 
as positronium is in electrodynamics. Neither charm nor bottom quarks 
are heavy enough to satisfy this requirement. And the top quark decays 
so rapidly that toponium has barely time to form. The binding of 
charmonia and bottomonia must be described non-perturbatively. Once the 
simplicity of a Coulombic bound state is foregone, having two heavy 
quarks to bind rather than a heavy and a light quark adds 
complications. Apart from the mass scale $m_Q$, set by the mass of the 
heavy quark, a quarkonium bound state involves three essential 
`small' (compared to $m_Q$) scales: $m_Q v$, the typical three-momentum of 
the constituents in the quarkonium rest frame or the inverse 
quarkonium size; $m_Q v^2$, the scale of binding energies and $\Lambda$. 
In a heavy-light meson, there is no other dimensionless ratio of scales 
besides $\Lambda/m_Q$. Although none of the 
non-perturbative properties of quarkonia are calculated 
in the approach described in the sequel, the multitude of low-energy 
scales entails more complicated power counting rules than those applicable 
to heavy-light mesons. One may envisage different power counting 
schemes, depending on the relation of the low energy scales. 

The success of non-relativistic potential models in describing static 
properties of the charmonium and bottomonium family suggests that 
these states are indeed non-relativistic and that $v^2$ could 
be used as a parameter for systematically expanding about the 
non-relativistic limit. With $v^2$ being small, quarkonium production 
involves 
two different time scales: the scale $1/m_Q$ on which a $Q\bar{Q}$ pair 
is produced\footnote{The scale $1/m_Q$ need not appear if the quarkonium
is produced through a weak interaction. Compare $B\to J/\psi X$ mediated 
by a $b\to c\bar{c} s$ transition with $B_c\to J/\psi X$ mediated by 
a $b\to c\bar{u} d$ transition. In the latter case, the $c\bar{c}$ pair is 
produced at distances of order of the $B_c$ radius.} and the scale 
$1/(m_Q v^2)$ on which the heavy quark pair binds into a quarkonium. 
Provided that the two stages of the production process can be 
separated and assuming that perturbation theory is valid at the 
scale $m_Q$, the heavy quark production part could be computed 
perturbatively; anything related to quarkonium formation could then 
be factorized into quarkonium-specific, but production process-independent, 
non-perturbative parameters.

A systematic realization of these ideas has been developed by Bodwin, 
Braaten and Lepage \cite{BBL}. It is based on an effective field theory, 
called 
non-relativistic QCD (NRQCD), combined with the methods of perturbative 
factorization and provides us with a tool to calculate inclusive quarkonium 
production cross sections as a double expansion in $\alpha_s$ and $v^2$, 
and to leading order in $\Lambda/m_Q$ in production processes with 
(light) hadrons in the initial state. This development is summarized 
in Sect.~2.

For phenomenology  the most important insight \cite{BF95} following 
from the NRQCD description of quarkonium production  
is that relativistic 
effects are very large in the production of ${}^3\!S_1$ 
states. Because $v^2$ is not very small, the radiation of gluons 
at late times in the production process, when the $Q\bar{Q}$ pair has 
already expanded to the quarkonium size, turns out to be favored as 
compared to the radiation from an almost point-like $Q\bar{Q}$ pair. 
As a consequence, the $Q\bar{Q}$ pair can remain in a color octet state 
at distances of order $1/m_Q$, a possibility that is ignored in the 
earlier color singlet model. The importance 
of these color octet contributions is supported by the large 
direct $J/\psi$ and $\psi'$ cross sections observed in $p\bar{p}$ 
collisions at the Tevatron \cite{tevatron}.

Subsequent to this initial success, almost all quarkonium production 
processes have been reconsidered in the light of NRQCD. The coverage 
of all production processes is beyond the scope of this presentation 
and I am restricting myself to an overview of the current status 
of inclusive quarkonium 
production at colliders and fixed target in Sects.~3 and 
4, respectively. A discussion of 
other interesting production processes, such as 
photoproduction, $e^+ e^-$ annihilation, $Z^0$ or $B$ decay, together 
with a (by now incomplete) list of references can be found in 
Refs.~\cite{BFY96,B,QQ}{}. There is also an increasing interest in 
polarization phenomena, as quarkonium polarization provides a 
calculable in NRQCD, and sometimes striking signature of color octet 
production. I believe that although the above choice of 
production processes is selective, it covers, on the one hand, 
the most dramatic new production mechanisms, and on the other hand 
illustrates the difficulties connected with a quantitative confirmation 
of the universality of long-distance parameters, as assumed in NRQCD. 
Especially for charmonium production, quarkonium binding effects 
that are at the center of tests of NRQCD factorization can rarely be 
exhibited in isolation from other QCD effects that reside in the 
short-distance part or are neglected, such as small-$x$ and soft 
gluon effects (colliders, photoproduction), BFKL-type situations 
(photoproduction, $z\to z_{\rm max}\approx 1$), 
higher-twist effects (fixed target, 
photoproduction), and parton-hadron duality ($B$ decay). Everything 
taken together makes for an intricate combination of QCD 
phenomena. Beyond any doubt, 
NRQCD is the correct theory for quarkonium systems in the heavy quark 
limit. Whether the charm quark mass is large enough to justify 
an expansion around this limit, will be decided by 
confronting predictions with experiments.
As of now, we are only beginning to assess theoretical 
uncertainties and to sort out those observables that eventually will 
stand as solid tests of NRQCD.
 
\section{Effective theory and factorization in quarkonium production}

\subsection{Non-Relativistic QCD}

Let me assume that indeed $m_Q\gg m_Q v, m_Q v^2,\Lambda$. In this situation 
of well-separated scales, it is intuitive that a quarkonium production 
cross section should factorize into the production cross section 
of a $Q\bar{Q}$ pair times the probability that this $Q\bar{Q}$ pair forms 
a specific quarkonium state. 

To make the scale separation explicit, one may think of integrating out 
all high momentum modes in the path integral down to a factorization 
scale $\mu$ in between $m_Q$ and $m_Q v$. The result would be an effective 
action, which is highly non-local on distances $1/m_Q$. But since the 
low momentum modes that dominate a quarkonium bound state can not 
resolve such small distances, this effective action could be expanded 
in an infinite series of local interactions. This expansion would realize 
an expansion in $v^2$; to all orders, it would be exactly equivalent to 
QCD. In practice, this procedure can be carried out only in very simplified 
situations. However, as long as the strong coupling $\alpha_s$ is small 
enough at the scale $\mu$, the effective Lagrangian can be constructed 
perturbatively to a specified accuracy.

First, one identifies the low-energy degrees of freedom. In the 
non-relativistic limit, intermediate states (in the sense of time-ordered 
perturbation theory) containing heavy quark pairs are suppressed and 
integrated out. The heavy quark field $Q$ splits into a two-spinor 
quark field $\psi$ and a two-spinor antiquark field $\chi$. The 
effective Lagrangian a priori consists of the most general Lagrangian, 
including non-renormalizable operators, consistent with the symmetries 
of QCD \footnote{The NRQCD Lagrangian is usually written in the heavy 
quark rest frame and is therefore constrained only by rotational symmetry. 
Of course, the result of any calculation is Lorentz invariant. The 
`hidden' full Lorentz symmetry constrains some of the coefficient 
functions in the Lagrangian.}. The 
coefficients of the operators are then tuned to reproduce QCD by 
comparing on-shell Green functions computed in QCD with those computed 
with the NRQCD Lagrangian. The desired accuracy and the values of 
$\alpha_s(\mu)$ and $v^2$ determine which operators have to be kept in 
NRQCD and to what loop-order the comparison has to be done. Note that 
since the `matching' is carried out at a scale much above the bound state 
scales, only scattering diagrams have to be computed. By construction, 
the NRQCD Lagrangian already has the same infrared 
behaviour as QCD. This implies 
in particular that the result of matching is independent of how 
the small scales $m_Q v$, $m_Q v^2$ and $\Lambda$ are related. 
However, their relation does have some consequences for which 
operators have to kept in NRQCD to achieve the desired accuracy.

Non-relativistic effective theory has first been introduced by 
Caswell and Lepage \cite{CL86} as a tool to manage bound state 
calculations in QED. This application is particularly transparent, 
as both the short-distance matching and the long-distance contributions 
can be calculated perturbatively. The main advantage of 
non-relativistic QED compared to the Bethe-Salpeter approach is that the 
physics above the scale $m_e$ is encoded once and forever in the 
effective Lagrangian. Because NRQED still contains two scales, 
$m_e\alpha$ and $m_e\alpha^2$, the scale separation is still 
not complete, but separating $m_e$ already entails great 
simplifications. See Ref.~\cite{KN} for a state-of-the-art calculation 
in NRQED.

The NRQCD Lagrangian takes the form
\begin{equation}
{\cal L}_{\rm NRQCD} = {\cal L}_2 + {\cal L}_4 +{\cal L}_{glue} 
+\ldots.
\end{equation} 
The light quark and gluon part of the QCD Lagrangian remains unaltered 
and is not indicated. 
The contribution to the effective Lagrangian that involves two heavy 
quark fields can be obtained at tree-level from a Foldy-Wouthuysen-Tani 
transformation, generalized to the non-abelian case (see e.g. 
Ref.~\cite{koerner}):
\begin{eqnarray}
\label{l2}
{\cal L}_2 &=& \psi^\dagger \left[iD_0+\frac{\vec{D}^2}{2 m_Q}\right]\psi + 
\frac{1}{8 m_Q^3}\psi^\dagger\vec{D}^4\psi + \frac{c_1}{2 m_Q}
\psi^\dagger\vec{\sigma}\cdot g\vec{B}\psi\nonumber\\
&&+\,\frac{c_2}{8 m_Q^2}\psi^\dagger(\vec{D}\cdot g\vec{E}-g\vec{E}\cdot 
\vec{D})\psi + \frac{c_3}{8 m_Q^2}\psi^\dagger(i\vec{D}\times g\vec{E}-
g\vec{E}\times i\vec{D})\psi\\
&&+\ldots + \,\mbox{charge-conjugated for the antiquark}\nonumber.
\end{eqnarray}
The leading term in square brackets describes a non-relativistic 
Schr\"odinger field theory. To reproduce the on-shell Green functions 
in QCD at order $v^2$, the four subsequent terms have to be included. 
Radiative corrections due to hard gluons shift their coefficients 
away from their tree level values $c_i=1$. The ellipses stand for 
higher order terms in $v^2$. To reproduce Green functions with 
$2 n$ external heavy quark fields, NRQCD must contain local operators 
with $2 n$ quark fields. In the following, only operators with 
four quark fields are of interest. The generic form of ${\cal L}_4$ is
\begin{eqnarray}
{\cal L}_4 &=& \sum_i \frac{d_i}{m_Q^2}(\psi^\dagger\kappa_i\chi) 
(\chi^\dagger\kappa_i^\prime\psi)\\
&&+\,\,\mbox{four quark scattering operators}\nonumber,
\end{eqnarray}
where $\kappa_i$ ($\kappa_i^\prime$) 
is a matrix in spin and colour indices and may 
also contain factors of spatial derivatives $\vec{D}/m_Q$ in 
case of higher-dimension operators. 
The coefficients $d_i$ of the annihilation operators are complex. 
Their imaginary parts describe the annihilation decay of 
quarkonium states \cite{BBL}. Finally, integrating out 
heavy quark loops leads to higher-dimension operators in 
gluon fields, ${\cal L}_{glue}$, the non-abelian analogue of the 
Euler-Heisenberg effective Lagrangian.

\begin{figure}[t]
   \vspace{-5cm}
   \centerline{\epsffile{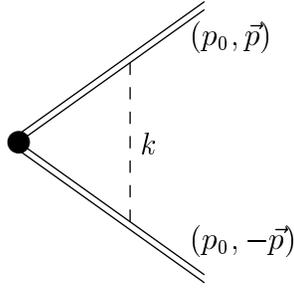}}
   \vspace*{-20cm}
\caption{\label{fig1}
Coulomb correction to point-like $\psi^\dagger\chi$ creation.}
\end{figure}

The NRQCD Lagrangian ${\cal L}_2+{\cal L}_{glue}$ coincides with 
the Lagrangian of heavy quark effective theory (HQET) \cite{hqet}. 
Nevertheless, the two effective theories are different, because 
their power counting schemes are different. Because $\Lambda$ is the 
only low-energy scale in heavy-light mesons, the importance of 
operators in the HQET Lagrangian is ordered strictly by dimension. 
Consequently, the kinetic energy operator $\psi^\dagger \vec{D}^2/
(2 m_Q)\psi$ is suppressed by $\Lambda/m_Q$ and the leading 
effective Lagrangian $\psi^\dagger iD_0\psi$ describes a static 
quark. In a non-relativistic bound state we expect $\vec{D}\sim 
m_Q v$, but $D_0\sim m_Q v^2$ and the kinetic term can not be 
neglected. A more compelling argument arises, if one begins to compute 
Green functions in the quark-antiquark sector of NRQCD. In the following  
I will implicitly assume the Coulomb gauge, which makes the physics 
of non-relativistic bound states most transparent. The amplitude  
for a $Q\bar{Q}$ pair created in a point and then interacting through 
exchange of a Coulomb gluon (see Fig.~\ref{fig1}), the `00'-component 
of the gluon propagator in this gauge, is given by 
\begin{equation}
\label{integral}
I = i C_F g_s^2\int\frac{d^4 k}{(2\pi)^4} \frac{1}{\vec{k}^2} 
\frac{1}{\left[p_0+k_0-\frac{(\vec{p}+\vec{k})^2}{2 m_Q}+i\epsilon\right]}
\frac{1}{\left[-p_0+k_0+\frac{(\vec{p}+\vec{k})^2}{2 m_Q}-i\epsilon\right]}.
\end{equation}
On-shell $p_0=\vec{p}^{\,2}/(2 m_Q)$. The integral is done most 
easily by closing the contour in the complex $k_0$-plane and 
picking up the residues of the enclosed poles. The poles are 
located at $k_0=p_0-(\vec{p}+\vec{k})^2/(2 m_Q)+i\epsilon$ and 
$k_0=-p_0+(\vec{p}+\vec{k})^2/(2 m_Q)-i\epsilon$, one on each side 
of the real axis. Then
\begin{equation}
\label{integral2}
I = C_F g_s^2\int\frac{d^3 k}{(2\pi)^3}\,
\frac{m_Q}{\vec{k}^2 (\vec{k}^2+2\vec{p}\cdot\vec{k}-i\epsilon)} = 
\frac{C_F\pi\alpha_s}{4}\,\frac{m_Q}{|\vec{p}|} + \,\mbox{imaginary}.
\end{equation}
The imaginary part is divergent and related to the 
Coulomb phase; the real part exhibits the 
well-known Coulomb divergence close to threshold. 

In the static approximation, the `kinetic term' in the propagators 
in (\ref{integral}) would be dropped and the on-shell condition 
reads $p_0=0$. In this limit the integration contour becomes pinched 
between the two poles and the integral becomes ill-defined. Thus, 
the kinetic term must be kept in the propagator to regulate 
the pinch-singularity. If one of the quarks were light, as 
in physical processes to which HQET applies, the static limit can be taken. 
Although the static propagator pole lies on the real axis, the contour 
can be deformed away from the pole. 

From (\ref{integral2}) it follows that the effective coupling for 
the exchange of Coulomb gluons 
is $\alpha_s m_Q/|\vec{p}|$. At small momenta 
$|\vec{p}|\sim m_Q\alpha_s$, Coulomb exchange can not be treated as a 
perturbation, even at weak coupling 
$\alpha_s$. The result of resumming Coulomb 
gluons is of course well-known and leads to Coulomb 
bound state poles in the quark Green function. On the other hand, 
HQET contains only strongly-coupled bound states. Note that, 
for the purpose of matching, it is sufficient to use free 
NRQCD propagators. Being of infrared origin, the Coulomb-enhanced 
terms cancel in the matching coefficients at every order. The integral in 
(\ref{integral2}) contains only one scale $|\vec{p}|=m_Q v$. Thus, 
its finite real part is dominated by gluons with momenta $m_Q v$ 
\footnote{However, they can not be thought of as on-shell intermediate 
states in time-ordered perturbation theory and in this sense 
they are not `dynamical'.}.

Because the leading order Lagrangian contains $m_Q$ in NRQCD, NRQCD 
does not lead to flavour symmetry. The non-perturbative properties 
of charmonia and bottomonia remain unrelated even in the non-relativistic 
limit. On the other hand, the NRQCD Lagrangian exhibits heavy-quark  
spin symmetry at leading order in $v^2$, which turns out to be 
useful to reduce the number of non-perturbative parameters that 
describe quarkonium binding.

I started out with quarkonium production, but the effective Lagrangian 
above does not yet allow quarkonia to be produced. One could not have 
expected that, because the short-distance part of a production process 
depends on the initial particles (hadrons, photons, $Z$ bosons ...) 
that initiate it. But for every production process, the separation 
of short and long distances can be performed in the same way as for the 
effective Lagrangian above. I will elaborate on this in the following 
subsection. 

At this point, let me pause for a discussion of the scales $m_Q$, 
$m_Q v$ and $m_Q v^2$ in charmonium and bottomonium systems. They 
are shown in Tab.~\ref{tab1}, together with those for positronium 
for comparison. The values of $v^2$ are based on potential models, 
which describe the spectrum of quarkonia reasonably accurately. 
For positronium 
$v\sim\alpha$. In a quantum field theoretic context, $v^2$ could 
be defined as the expectation value of a derivative operator that 
scales like $v^2$ according to the scaling rules to be described later. 
In general, $v^2$ will be a complicated function of $m_Q$ and 
$\Lambda$. In the limit 
$m_Q\to\infty$, the binding becomes Coulombic and 
$v\sim1/\ln (m_Q/\Lambda)$. Neither charmonia nor bottomonia are 
Coulombic, because the energy scale $m_Q v^2$ is of order $\Lambda$ for 
both quarkonium families. The same conclusion can be obtained 
from the observation that with $r\sim 1/(m_Q v)$, the 
contribution of the linear term in the Cornell potential 
\cite{cornell}
\begin{equation}
V(r)=-\frac{C_F\alpha_s(m_Q)}{r} + a^2 r
\end{equation}
\begin{table}[t]
\addtolength{\arraycolsep}{0.2cm}
\renewcommand{\arraystretch}{1.25}
$$
\begin{array}{c|ccc}
\hline\hline
 & c\bar{c} & b\bar{b} & e^+ e^-  \\ 
\hline
m_Q        & 1.5\,\mbox{GeV} & 5\,\mbox{GeV} & 0.5\,\mbox{MeV}\\
m_Q v      & 750\,\mbox{MeV} & 1.4\,\mbox{GeV} & 3.7\,\mbox{keV} \\
m_Q v^2    & 400\,\mbox{MeV} & 400\,\mbox{MeV} & 25\,\mbox{eV} \\ 
\hline
v^2        & 0.25 & 0.08 & 5\cdot 10^{-5} \\ 
\alpha(m_Q)/\pi & 0.1 & 0.07 & 2\cdot 10^{-3} \\
\hline\hline
\end{array}
$$
\caption{\label{tab1}
Scales in onium systems and the expansion parameters of the 
non-relativistic approximation.}
\end{table}
\noindent is non-negligible with respect to the Coulomb term. (The string 
tension is $a\approx 430\,$MeV and $C_F=4/3$.) Let me stress that 
NRQCD does not rely in any way on a Coulombic system. It is 
enough that $m_Q$ is large compared to all bound state scales. 
The values in Tab.~\ref{tab1} should be understood for the ground 
state of each family. Excited states are more non-relativistic and 
at the same time even less Coulombic. It is possible that the 
hierarchy of $m_Q v$ and $m_Q v^2$ with respect to $\Lambda$ 
changes as one 
considers higher excited quarkonium states. Different power counting 
rules would then apply to different members of the onium family. Comparing 
the different onium systems, note that relativistic corrections  
(governed by the parameter $v^2$) are exceedingly small for positronium, 
comparable to radiative corrections (governed by the parameter 
$\alpha_s(m_Q)/\pi$) for bottomonium and definitively large for 
charmonium. 
 
\subsection{Factorization and matching}
\label{sectfact}

The factorization of a quarkonium production process, 
suggested in Ref.~\cite{BBL}{}, begins with the observation 
that the creation of the $Q\bar{Q}$ pair requires an energy larger 
than $2 m_Q$ and therefore some of the propagators of the diagram 
are off-shell by at least $m_Q^2$, much larger than the typical 
off-shellness in a quarkonium bound state. These propagators can be 
`contracted to a point', and the remainder of the diagram, sensitive 
to distances larger than $1/m_Q$, is then associated 
with an operator matrix element in NRQCD. This is true, provided 
the production process is inclusive, 
\begin{equation}
\label{eq}
A+B\to \psi(P,\lambda) + X,
\end{equation}
where $X$ denotes light hadrons and $\lambda$ the quarkonium polarization 
state. If it is not inclusive, the process is sensitive to the details 
of the hadronic final state 
and its long-distance contributions can not be absorbed 
in quarkonium-specific objects alone. 

\begin{figure}[t]
   \vspace{-6cm}
   \centerline{\epsffile{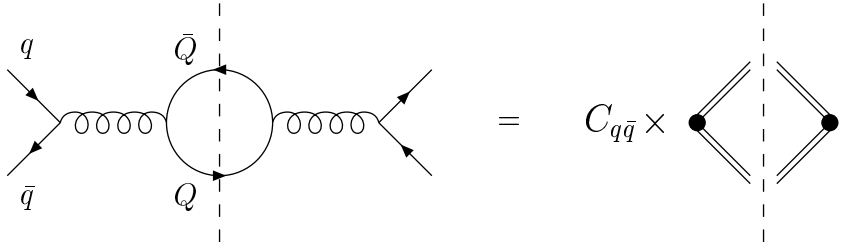}}
   \vspace*{-20cm}
\caption{\label{fig2} Trivial example of factorization. 
}\end{figure}

The short-distance coefficients can be computed by replacing 
the quarkonium in the final state by a perturbative $Q\bar{Q}$ state 
at small relative momentum, because the short-distance coefficient 
by construction does not depend on those effects that bind such 
a quark state into the quarkonium. Hence one can (and should) 
use on-shell quark states and, instead of (\ref{eq}), 
one considers
\begin{equation}
\label{eq2}
A+B\to Q(p) \bar{Q}(\bar{p}) +\, \mbox{gluons, light quarks},
\end{equation}
where
\begin{equation}
p=\left(\sqrt{m_Q^2+\vec{q}_1^{\,2}},\vec{q}_1\right)
\qquad\quad
\bar{p}=\left(\sqrt{m_Q^2+\vec{q}_2^{\,2}},\vec{q}_2\right)
\end{equation}
in the frame where the $Q\bar{Q}$ pair is almost at rest. (We may 
think of this frame as the quarkonium rest frame.) The amplitude 
squared is then expanded in the small quantities $\vec{q}_1$, 
$\vec{q}_2$ and the external heavy quark spinors are expressed in 
terms of two-component spinors $\xi$ and $\eta$. For a quarkonium 
not at rest, one applies a Lorentz boost to the vectors 
$p$ and $\bar{p}$. For the case $\vec{q}_2=-\vec{q}_1$ explicit 
formulae for the reduction of the amplitude to rest frame 
two-spinors can be found in Ref.~\cite{bc}{}. The simplest example 
of tree-level matching is shown in Fig.~\ref{fig2}. Evaluating 
the cut diagram, summing over all polarizations results in 
\begin{equation}
\label{example}
\frac{16\pi^3\alpha_s^2}{27 (2 m_Q)^3}\,\delta(\hat{s}-4 m_Q^2)\,
\times\,\eta^\dagger\sigma^i T^a\xi\,\xi^\dagger\sigma^i T^a\eta + 
O({\vec{q}_1}^{\,2},{\vec{q}_2}^{\,2},\vec{q}_1\cdot\vec{q}_2),
\end{equation}
where $\hat{s}$ is the cms energy of the $q\bar{q}$ pair. The spinor 
product can be identified with the tree-level evaluation of 
the matrix element
\begin{equation}
\label{me}
\langle {\cal O}_8^\psi({}^3\!S_1)\rangle\equiv 
\sum_{X,\lambda}\langle 0|\chi^\dagger\sigma^i T^a\psi
|\psi(\lambda)+X\rangle\langle\psi(\lambda)+X|\psi^\dagger 
\sigma^i T^a\chi|0\rangle
\end{equation}
where $\psi$ is again replaced by a $Q\bar{Q}$ pair at small 
relative momentum. Eq.~(\ref{me}) displays the  structure 
of a quarkonium production matrix element. In general, 
the $Q\bar{Q}$ pair can be in various color and spin states. 
Evidently, in our example, the $Q\bar{Q}$ pair must 
be in a color octet and spin-one state as expressed by the 
product $\sigma^i T^a$ in (\ref{me}). Further terms in the 
expansion in $\vec{q}_1$ and $\vec{q}_2$ can be associated 
with operators similar in form, but with derivatives.  It is 
natural to combine these into `relative' and `cms' derivatives, 
such that the following identifications can be made
\begin{eqnarray}
&& (q_1-q_2)_k\,\,\xi^\dagger\eta\quad\to\quad \psi^\dagger
\left(i\stackrel{\leftrightarrow}{D}_k\right)\chi
\nonumber\\
&& (q_1+q_2)_k\,\,\xi^\dagger\eta\quad\to\quad i D_k\left(\psi^\dagger 
\chi\right).
\end{eqnarray}
Operators with one relative derivative on each fermion bilinear 
can be decomposed as ${}^3\!P_J$ with $J=0,1,2$. It is 
important that the quantum numbers of the $Q\bar{Q}$ pair in the 
operator need not coincide with the quantum numbers of the 
quarkonium $\psi$, because they refer to the $Q\bar{Q}$ state 
at a time $\tau\sim 1/m_Q$ (in the quarkonium rest frame), long 
before the quarkonium is formed. In between gluons with 
energies less than $m_Q$ can be emitted (so that $X$ becomes 
non-trivial) and change the color and 
spin of the $Q\bar{Q}$ pair. By the nature of factorization, these 
low energy gluons have to be and are included in the definition of the 
non-perturbative parameters above. 

Operators with cms derivatives are usually neglected, because they 
are suppressed in $v^2$ according to the power counting rules discussed 
below\footnote{For quarkonium decays such operators were introduced 
by Mannel and Schuler \cite{MS}.}. These operators are important to 
resolve an ambiguity that arises at leading order in $v^2$ and which 
is apparent in (\ref{example}): the phase space restrictions ($\hat{s}=
4 m_Q^2$) are expressed in terms of partonic variables, such as the 
quark mass, and do not reflect the physical phase space for a 
quarkonium state with mass different from $4 m_Q^2$. It must be like this, 
because the phase space is part of the short-distance coefficient and 
therefore can depend only on short-distance parameters. It is also 
consistent, because $M_\psi-2 m_Q\sim m_Q v^2$ and the mass difference 
can be neglected at leading order in $v^2$. This negligence is 
certainly unjustified, if an observable is sensitive to the kinematic 
boundaries of phase space and leads to large ambiguities even 
in fully inclusive (and $p_t$-integrated) hadroproduction cross 
sections at high energies, because of the steep rise of the gluon 
distribution at small $x$. In the example of Fig.~\ref{fig2} 
$\vec{q}_1+\vec{q}_2$ enters the phase-space delta-function. 
Expansion in $\vec{q}_1+\vec{q}_2$ 
leads to a series of higher-dimension operators with 
increasingly singular coefficients, schematically written as 
\begin{equation}
\sum_n c_n\,\delta^{(n)}(\hat{s}-4 m_Q^2)\,(\chi^\dagger\sigma^i T^a
\psi)\,(\vec{D})^n\,(\psi^\dagger\sigma^i T^a\chi),
\end{equation}
where the superscript on the delta-function denotes derivatives. 
The resummation of this series leads to a (non-perturbative) 
distribution function \cite{BRW} with support properties such that 
its convolution with the short-distance coefficient reproduces 
the physical phase space constraints\footnote{In the context 
of quarkonium decays, the breakdown of the NRQCD expansion near 
boundaries of phase space is discussed in Refs.~\cite{MW,RW}{}.}. 
We will meet a particular application of this in Sect.~\ref{crosstev}.

So far I have discussed only tree-level matching. Factorization 
becomes non-trivial, when one computes radiative corrections, 
for instance to the leading-order $q\bar{q}$ annihilation 
process in Fig.~\ref{fig2}:
\begin{equation}
\label{radiative}
q+\bar{q} \to Q(p) \bar{Q}(\bar{p}) + g.
\end{equation}
When the $Q\bar{Q}$ pair is in a $P$-wave state, the cross 
section is infrared divergent, when the gluon is soft\cite{barbieri}. 
(I assume that collinear singularities from emission off the 
initial quarks have already been absorbed into redefined partons.) 
For a long time, this infrared divergence has shed doubt on the perturbative 
calculability of $P$-wave quarkonium decay or production. Within the 
NRQCD approach, this problem finds its natural solution: \cite{bbl2} 
the infrared divergence indicates that the soft gluon emission is sensitive 
to bound state scales and therefore should be factored in a NRQCD matrix 
element. Since before gluon emission, the $Q\bar{Q}$ pair 
is in a color octet, spin-one state, the only candidate is 
$\langle {\cal O}_8^\psi({}^3\!S_1)\rangle$. Indeed, since this 
matrix element appeared at leading order (Fig.~\ref{fig2} 
and (\ref{example})), a complete matching calculation includes 
the $\alpha_s$ correction to this matrix element, computed 
in a perturbative $Q\bar{Q}$ state at small relative momentum. 
This contribution is both ultraviolet and infrared divergent. 
The ultraviolet divergence can be absorbed into a 
renormalization of $\langle {\cal O}_8^\psi({}^3\!S_1)\rangle$, 
which becomes explicitly factorization scale dependent. 
The infrared divergence cancels the infrared 
divergence in the calculation 
of the process (\ref{radiative}). The short-distance coefficient 
of the $P$-wave matrix element is now infrared finite, but contains 
a scale-dependent $\ln(m_Q/\mu)$. The scale-dependence cancels 
with the scale-dependence of the octet matrix element 
$\langle {\cal O}_8^\psi({}^3\!S_1)\rangle(\mu)$. 

Note that before the advent of NRQCD, the infrared 
logarithm $\ln m_Q/\lambda$ 
was sometimes treated as an adjustable non-perturbative parameter, see 
for instance Ref.~\cite{SCH94}{}. 
When this is done consistently in all production processes, this 
procedure is fully equivalent to taking into account the 
color octet contribution $\langle {\cal O}_8^\psi({}^3\!S_1)\rangle$ 
in NRQCD. It is for this reason that the leading color octet contributions 
in NRQCD for $P$-wave production do not lead to significant 
differences compared to the color singlet model \cite{csm} 
at next-to-leading order. Ironically, color octet 
contributions turn out to be more 
important for $S$-wave quarkonia, precisely because they are 
suppressed in $v^2$ and not intertwined with the leading color singlet 
term through (logarithmic) operator mixing. 

\begin{figure}[t]
   \vspace{-4.5cm}
   \centerline{\epsffile{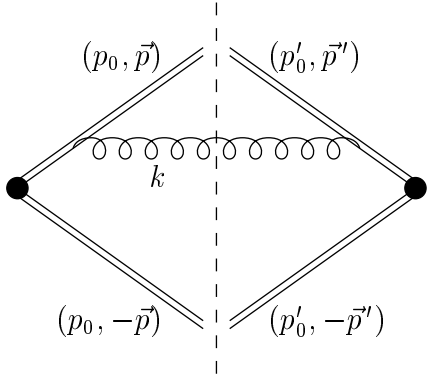}}
   \vspace*{-19.5cm}
\caption{\label{fig3} Contribution to 
mixing of $\langle {\cal O}_8^\psi({}^3\!S_1)\rangle$ into  
$\langle {\cal O}_1^\psi({}^3\!P_0)\rangle$.}
\end{figure}

It is quite instructive to go into some details of the NRQCD part 
of the matching calculation. As an example, let me consider the  
one-loop mixing of $\langle {\cal O}_8^\psi({}^3\!S_1)\rangle$ into 
$\langle {\cal O}_1^\psi({}^3\!P_0)\rangle$. To get a non-zero 
contribution, a transverse gluon must be cut. The dimensions work 
out correctly, because a transverse gluon couples proportional to 
$\vec{p}/m_Q$. One of the diagrams is shown in Fig.~\ref{fig3}; it 
leads to an integral 
\begin{equation}
\label{integral3}
I = \int\frac{d^3 k}{(2\pi)^3} \frac{1}{2|\vec{k}|} \,
\frac{1}{m_Q^2}
\frac{\vec{p}\cdot\vec{p}^{\,\prime}-
(\vec{p}\cdot k)(\vec{p}^{\,\prime}\cdot k)/\vec{k}^{\,2}}
{\left[|\vec{k}|-\frac{2\vec{k}\cdot\vec{p}+\vec{k}^{\,2}}
{2 m_Q}+i\epsilon\right]
\left[|\vec{k}|-\frac{2\vec{k}\cdot\vec{p}^{\,\prime}+\vec{k}^{\,2}}
{2 m_Q}+i\epsilon\right]}.
\end{equation}
The integral is divergent and has to be regularized. One would like 
to use dimensional regularization, because it makes matching calculations 
particularly simple. Dimensional regularization is fine, as long as the 
effective theory contains only one low-energy scale, 
such as HQET. Then all integrals 
in the effective theory are scaleless and vanish identically. 
NRQCD integrals such as (\ref{integral3}) are not of this type. 

To see how such integrals are evaluated, let me consider the following 
simplified version of (\ref{integral3}), both with a cut-off and dimensional 
regularization:
\begin{eqnarray}
J_{\Lambda_c} = \int\limits_\lambda^{\Lambda_c}\frac{dk}{k} 
\frac{m_Q^2}{(m_Q+p+k)^2}\nonumber\\
J_{d} = \mu^{-\epsilon}
\int\limits_\lambda^{\infty}\frac{dk}{k^{1-\epsilon}} 
\frac{m_Q^2}{(m_Q+p+k)^2}.
\end{eqnarray}
I will only consider the logarithmically divergent and finite contributions 
in the limit of small infrared cut-off $\lambda$. Because NRQCD is an 
effective theory below the scale $m_Q$, the ultraviolet cut-off must be 
chosen such that $p\ll \Lambda_c\ll m_Q$. Therefore, taking the integral 
$J_c$ and expanding it in $\Lambda_c/m_Q$ and $p/m_Q$, one gets
\begin{equation}
\label{cutoff}
J_{\Lambda_c} = \left(1-\frac{2 p}{m_Q}\right)
\ln\frac{\Lambda_c}{\lambda}-\frac{2\Lambda_c}{m_Q} + \ldots.
\end{equation}
The same result would have been obtained, if the integrand had been first  
expanded in $p/m_Q$ and $k/m_Q$, which are both small for $k<\Lambda_c$. 
Note the logarithmic term in the cut-off $\Lambda_c$, which would 
correspond to mixing of $\langle {\cal O}_8^\psi({}^3\!S_1)\rangle$ into  
$\langle {\cal O}_1^\psi({}^3\!P_0)\rangle$ in the real case. The 
dimensionally regulated integral, in the limit $\epsilon\to 0$,  evaluates to 
\begin{equation}
\label{dim1}
J_d=\frac{m_Q^2}{(m_Q+p)^2}\left(\ln\frac{m_Q+p}{\lambda}-1\right)
\end{equation}
and seems to have no ultraviolet divergence. The problem here is that 
dimensional regularization does not know about the physical 
requirement $p\ll \Lambda_c\ll m_Q$, that expresses that the factorization 
scale is smaller than $m_Q$, and treats the cut-off as if it were 
larger than $m_Q$. Indeed, up to power-like cut-off dependence 
(\ref{dim1}) coincides with the result for $J_{\Lambda_c}$ evaluated 
for $\Lambda_c\gg m_Q$. We can force dimensional regularization to 
treat $m_Q$ larger than all other scales by expanding the 
integral in $m_Q$ before integration. The result is 
\begin{equation}
\label{dim2}
J_d=\left(1-\frac{2 p}{m_Q}\right) \left(\frac{1}{\epsilon}-
\ln\frac{\lambda}{\mu}\right) + \ldots,
\end{equation}
in agreement with (\ref{cutoff}) up to power-like terms in the 
cut-off which are 
always zero in dimensional regularization. Thus, 
in dimensional regularization 
one must expand the integrand before integration, while with a cut-off 
integrating before or after expansion yields the same result, provided 
$\Lambda\ll m_Q$. The calculation may be simplified even further, when 
the infrared 
divergence is also regulated dimensionally. After expansion of the 
integrand, all integrals are scaleless and vanish in dimensional 
regularization. This technique has been used in Ref.~\cite{manohar} 
to obtain the NRQCD Lagrangian up to order $\alpha_s/m_Q^3$ in the 
single heavy-quark sector.

Expanding the integrand works for matching calculations 
to all loops in the single heavy-quark sector, but fails in the 
quark-antiquark sector. Indeed, we know already from 
(\ref{integral2}) that some matrix elements in the effective theory 
must be non-vanishing, so that the Coulomb divergence cancels 
in the matching. Integrands containing Coulomb singularities can 
not be expanded in $m_Q$ before integration over $k_0$. While at one-loop  
it is easy to separate the Coulomb contributions explicitly, 
a general matching scheme based 
on dimensional regularization that would treat diagrams with mixed 
Coulomb and transverse gluon exchanges has not yet been devised. 
Since the problem does not appear 
in matching calculations with a cut-off\footnote{NRQED calculations 
are naturally done in such a scheme, see e.g. Ref.~\cite{KN}{}. Otherwise, 
the bound state diagrams would also have to be evaluated in dimensional 
regularization.}, 
it is not related with the effective NRQCD Lagrangian per se. 
See, however, Ref.~\cite{ira} for an alternate view of the problem. 

\begin{figure}[t]
   \vspace{0cm}
   \epsfysize=7cm
   \epsfxsize=10cm
   \centerline{\epsffile{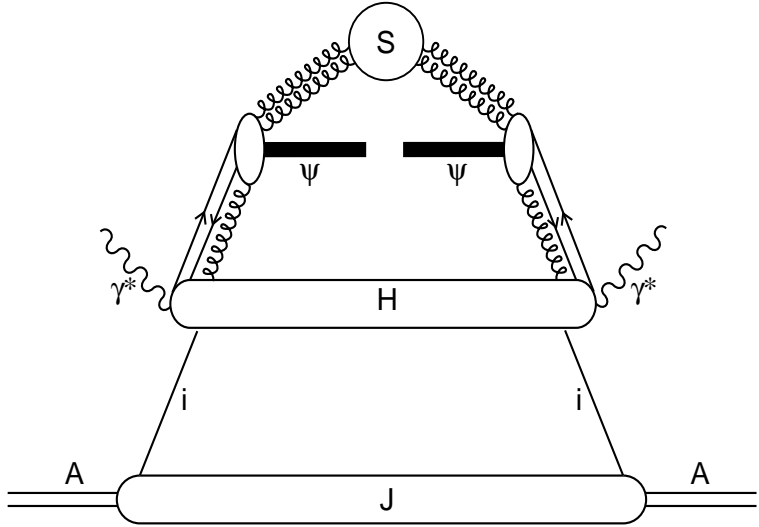}}
   \vspace*{0cm}
\caption{\label{fig4} Diagrammatic representation of factorization in 
$\gamma^*+A\to \psi+X$. For the cross section the diagram has to be 
cut.} 
\end{figure}

Summarizing the discussion of this subsection, the differential 
quarkonium production cross section in a hadron-hadron collision, 
$A+B\to \psi(P,\lambda) + X$, can be 
factorized as
\begin{equation}
\label{factform}
d\sigma = \sum_{i,j}\int dx_1 dx_2\,f_{i/A}(x_1) f_{j/B}(x_2)
\sum_n d\hat{\sigma}_{i+j\to Q\bar{Q}[n]+X}\,
\langle {\cal O}^{\psi(\lambda)}_n\rangle.
\end{equation}
This equation is diagrammatically represented in Fig.~\ref{fig4}, where 
I have replaced hadron $B$ by a virtual photon for graphical 
simplicity. Each factor in (\ref{factform}) corresponds to a subgraph 
in Fig.~\ref{fig4} and a momentum scale that dominates this subgraph. 
The amplitude squared for $A\to i+\,\mbox{remnant}\,J$ is given by the 
parton distribution $f_{i/A}(x_1)$. The typical virtualities in this 
subgraph are of order $\Lambda^2$. Parton $i$ then participates in a 
hard collision $H$, in which a $Q\bar{Q}$ pair in a certain state $n$ 
is produced. This involves energies of order $m_Q$. The $Q\bar{Q}$ pair 
(and, in general, additional gluons) connects to the quarkonium subgraph, 
represented by the NRQCD matrix element $\langle 
{\cal O}^{\psi(\lambda)}_n\rangle$, 
which contains all soft lines that are sensitive to the 
bound state scale $m_Q v^2$. After factorizing initial state 
singularities into redefined parton distributions and performing 
the NRQCD matching as described above, the short-distance 
cross section $d\hat{\sigma}_{i+j\to Q\bar{Q}[n]+X}$ is infrared finite, 
but depends on the collinear and NRQCD factorization scale. The scale 
dependence cancels in the product (\ref{factform}). 

Note that the factorization formula above implies complete factorization 
between final and initial state; all lines connecting both run through 
the hard part of the diagram. The cancellation of soft gluons, that 
could connect the quarkonium and remnant jet part, is believed, but 
not yet really proven, to hold 
perturbatively for inclusive quarkonium production. This is quite 
intuitive, if the quarkonium is scattered at large transverse momentum 
with respect to the colliding hadrons. It holds also for the total 
quarkonium production cross section even though the bulk cross 
section is from quarkonia at small transverse momentum. This is 
intuitive only in the Coulombic limit $m_Q v^2\gg \Lambda$, in 
which quarkonium formation has terminated long before the time-scale 
of a second interaction with the remnant. Since perturbative 
matching is insensitive to the relation of low energy scales, 
factorization should thus hold in perturbation theory. The calculation 
of next-to-leading corrections to total $P$-wave production cross 
sections \cite{MP} confirms factorization at the one-loop 
level. The mathematical statement of factorization is thus that 
all corrections to (\ref{factform}) scale as some power of $1/m_Q$,  
or another kinematic scale larger than $m_Q$, as $m_Q\to \infty$. 
(Recall that relativistic and radiative corrections 
scale as $1/\ln m_Q$ in this limit.) Corrections should then be 
suppressed as $\Lambda$ divided 
by potentially any other scale in the process. Since 
$m_Q v^2\sim \Lambda$ for charmonium and bottomonium, one may 
expect large `higher-twist' corrections in fixed-target 
collisions, when the heavy quark-antiquark pair 
moves parallel with a remnant jet 
and remains in its hadronization region over a time $1/\Lambda$ 
in the quarkonium rest frame. 
Even if higher-twist effects scale only as $\Lambda^2/m_c^2$ for 
charmonium, they can be expected to be non-negligible for total 
cross sections.

\subsection{Power counting (velocity scaling)}

The factorization formula (\ref{factform}) contains an infinite 
series of non-perturbative production matrix elements 
$\langle {\cal O}^{\psi(\lambda)}_n\rangle$ and would be useless, 
if it could not be truncated after a finite number of terms. A 
first indication comes from the dimension of the operators, since any 
power of $1/m_Q$ in the coefficient of an operator can be 
compensated only by one of the low-energy scales $m_Q v$,  
$m_Q v^2$ or $\Lambda$. However, the matrix elements can have 
additional suppressions beyond their dimension, because of the 
particular structure of a non-relativistic bound state. 

The standard power counting (or `velocity scaling') rules are 
due to Ref.~\cite{vel}{}. For example, heavy quark fields scale as 
$(m_Q v)^{3/2}$, because $\int d^3 r\,\psi^\dagger\psi$ counts 
heavy quark number and $r\sim 1/|\vec{p}|\sim 1/(m_Q v)$. The Virial 
theorem relates the potential to the kinetic energy $E\sim m_Q v^2$ 
and leads to $g\vec{E}\sim m_Q^2 v^3$ for the electric field inside 
a quarkonium. From the equation of motion for the vector potential, 
$g\vec{B}\sim m_Q^2 v^4$ follows for the magnetic field. As a 
consequence the `dipole interaction' $\psi^\dagger (g\vec{A}\cdot 
\vec{\partial}/m_Q^2)\psi$ and `magnetic interaction' 
$\psi^\dagger (g\vec{B}\cdot\vec{\sigma}/(2 m_Q))\psi$ in the 
NRQCD Lagrangian both scale 
as $v^2$ relative to the Coulomb interaction. 

It follows that a matrix element such as 
$\langle {\cal O}_1^{J/\psi}({}^3\!S_1)\rangle$ scales as $v^3$, 
because a $Q\bar{Q}$ pair in a color singlet ${}^3\!S_1$ state 
overlaps with the leading Fock state wavefunction of a $J/\psi$ 
without need of soft gluon emission\footnote{Up to corrections 
in $v^2$, this matrix element coincides with the wavefunction 
at the origin squared. Keeping only this contribution, NRQCD reproduces 
the color singlet model for $S$-wave quarkonia.}. In the following 
the scaling of all matrix elements will be considered relative to 
$\langle {\cal O}_1^{J/\psi}({}^3\!S_1)\rangle$, that is, I put 
$\langle {\cal O}_1^{J/\psi}({}^3\!S_1)\rangle\sim 1$. To 
estimate the scaling of other matrix elements, the multipole 
suppression of gluon emission with momentum $\vec{k}\sim m_Q v^2 
\ll \vec{p} \sim m_Q v$ has to be taken account as a consequence 
of the two low energy scales present in NRQCD. (In the context of 
NRQED, the separation of photons with momenta $|\vec{k}|\sim m_e v$ 
and $|\vec{k}|\sim m_e v^2$ and the power counting for the two 
momentum regions has been analyzed in detail by Labelle 
\cite{labelle}.) 

Consider the $P$-wave matrix element $\langle {\cal O}_8^{J/\psi}
({}^3\!P_0)\rangle$ in a $J/\psi$ state. A non-zero overlap requires 
the emission of gluons into the final state. For a single gluon 
emission, one such diagram looks exactly like Fig.~\ref{fig3}. The 
$Q\bar{Q}$ vertex carries a derivative for the $P$-wave operator 
and the transverse gluon also couples proportional to $\vec{p}/m_Q$ 
from the dipole interaction term above. Thus $\langle {\cal O}_8^{J/\psi}
({}^3\!P_0)\rangle$ scales as $p^4/m_Q^4\sim v^4$ relative to 
$\langle {\cal O}_1^{J/\psi}({}^3\!S_1)\rangle$, to which diagrams 
with out radiation would contribute. 

Consider now $\langle {\cal O}_8^{J/\psi}({}^1\!S_0)\rangle$, which 
requires a spin-flip transition. As the chromomagnetic interaction 
is proportional to the gluon momentum, a diagram such as Fig.~\ref{fig3} 
leads to (cf. (\ref{integral3})) 
\begin{eqnarray}
\label{m1}
I &=& \frac{\alpha_s}{4 m_Q^2}\int\limits^{\,\Lambda_c}
\frac{d^3 k}{(2\pi)^3} 
\frac{1}{2|\vec{k}|} \,
\frac{P_{ij}\,(\vec{k}\times\vec{\sigma})_i\,(\vec{k}\times\vec{\sigma})_j}
{\left[p_0+|\vec{k}|-\frac{(\vec{p}+\vec{k})^2}
{2 m_Q}+i\epsilon\right]
\left[p_0^\prime+|\vec{k}|-\frac{(\vec{p}^{\,\prime}+\vec{k})^2}
{2 m_Q}+i\epsilon\right]}
\nonumber\\
&\sim& \alpha_s\frac{\Lambda_c^2}{m_Q^2} + \,\mbox{magnetic dipole 
contribution,}
\end{eqnarray}
where $P_{ij}=\delta_{ij}+k_i k_j/\vec{k}^{\,2}$. In the second line 
I have (schematically) separated the contribution from the region 
$|\vec{k}|\sim m_Q v$. In this region the $|\vec{k}|$-terms in the 
quark propagators dominate, the integrand becomes independent of the 
bound state structure and results in a pure (power-like) cut-off term. 
The contribution from $|\vec{k}|\sim m_Q v^2$ is denoted as `magnetic 
dipole' and scales as 
\begin{equation}
\frac{\alpha_s}{m_Q^2}\,|\vec{k}|^2\sim v^2\lambda^2\sim v^4,
\end{equation}
where for $|\vec{k}|\sim m_Q v^2$ the coupling $\alpha_s$ must be counted 
as 1. $\lambda \equiv k/p \sim v$ corresponds to the ratio of the size 
of the quarkonium and the wavelength of the emitted gluon and provides 
the expansion parameter for the multipole expansion. In general, 
once gluons of momentum $m_Q v$ are separated, the interaction vertices 
of NRQCD can be multipole-expanded. This is true beyond perturbation 
theory and justifies the single-gluon approximation which I used 
for illustration. Thus $\langle {\cal O}_8^{J/\psi}({}^1\!S_0)\rangle$ 
scales as $v^4$, if the cut-off term in (\ref{m1}) could 
be discarded. If one identifies $\Lambda_c\sim m_Q v$ and $\alpha_s$ at 
the scale $m_Q v$ with $v$, as suggested by a Coulombic limit, the 
cut-off term scales as \cite{err,schuler} $v^3$. 
Being pure cut-off, however, it 
should not be taken into account to estimate the scaling of low-energy 
matrix elements. It would 
be cancelled with a contribution to the coefficient function, 
if it were evaluated with the same cut-off. In dimensional regularization, 
the contribution from $|\vec{k}|\sim m_Q v$ gives a tadpole-like integral 
and would be set to zero naturally. 

The multipole suppression is effective as long as $m_Q v\gg
\Lambda \sim m_Q v^2$ holds. If, on the other hand, 
$m_Q v\sim \Lambda\gg m_Q v^2$, 
the typical momenta of (soft) gluons would be $\Lambda$ and the 
multipole expansion would not work, as $\lambda=1$. We can keep $\lambda$ 
as a free parameter and conceive an intermediate case for 
$J/\psi$ or, in particular, $\psi'$. The power counting for the 
most important $Q\bar{Q}$ states is summarized in Tab.~\ref{tab2} and 
3 for $S$- and $P$-wave quarkonia, respectively. 
For each $Q\bar{Q}$ state, there exist $v^2$ corrections 
due to operators with two and more 
derivatives on a single bilinear of fermion fields. Since their 
coefficient functions are not enhanced by fewer powers of $\alpha_s$ 
compared to the leading operator in each channel, I will not discuss 
them further. Like operators with cms derivatives, these operators can 
be important in specific kinematic regions.  

\begin{table}[t]
\addtolength{\arraycolsep}{0.1cm}
\renewcommand{\arraystretch}{1.25}
$$
\begin{array}{c|cccc}
\hline\hline
 & \langle {\cal O}_1^{\psi}({}^3\!S_1)\rangle & 
   \langle {\cal O}_8^{\psi}({}^3\!S_1)\rangle & 
   \langle {\cal O}_8^{\psi}({}^1\!S_0)\rangle &
   \langle {\cal O}_8^{\psi}({}^3\!P_0)\rangle \\ 
\hline
\mbox{NRQCD}  & 1 & v^4 & v^2\lambda^2 & v^4 \\
\mbox{CEM}    & 1 & 1   & 1            & v^2 \\
\hline\hline
\end{array}
$$
\caption{\label{tab2}
Velocity scaling in NRQCD and the color evaporation model (CEM) 
relative to 
$\langle {\cal O}_1^{\psi}({}^3\!S_1)\rangle$ for 
$S$-wave quarkonia. 
The `standard' velocity counting is recovered for $\lambda=v$.}
\end{table}
\begin{table}[t]
\addtolength{\arraycolsep}{-0.1cm}
\renewcommand{\arraystretch}{1.25}
$$
\begin{array}{c|cccccc}
\hline\hline
& \langle {\cal O}_1^{\psi}({}^3\!P_J)\rangle &  
\langle {\cal O}_8^{\psi}({}^3\!S_1)\rangle & 
\langle {\cal O}_8^{\psi}({}^3\!P_{J'})\rangle & 
\langle {\cal O}_8^{\psi}({}^1\!P_1)\rangle &
\langle {\cal O}_1^{\psi}({}^3\!S_1)\rangle &
\langle {\cal O}_8^{\psi}({}^3\!D_{J''})\rangle 
\\ 
\hline
\mbox{NRQCD}  & v^2 & v^2 & v^6 & v^4\lambda^2 & v^6 & v^6\\
\mbox{CEM}    & v^2 & 1   & v^2 & v^2          & 1   & v^4\\
\hline\hline
\end{array}
$$
\caption{\label{tab3}
Velocity scaling in NRQCD and the color evaporation model (CEM) 
relative to 
$\langle {\cal O}_1^{\psi}({}^3\!S_1)\rangle$ for 
$P$-wave quarkonia with total angular momentum $J$.  
The `standard' velocity counting is recovered for $\lambda=v$.}
\end{table}

Before closing this section, I would like to mention the color 
evaporation model \cite{cem} (CEM), 
the only remaining potential competitor 
of NRQCD \footnote{As alluded to earlier, the color singlet model 
has been swallowed by NRQCD and lost its justification hence after.}. 
In the CEM, the NRQCD expansion on the right hand side of 
(\ref{factform}) becomes replaced by an average, 
\begin{equation}
\label{cem}
\sum_n d\hat{\sigma}_{i+j\to Q\bar{Q}[n]+X}\,
\langle {\cal O}^{\psi(\lambda)}_n\rangle
\to
f_\psi\int\limits_{2 m_Q}^{2 m_{Qq}} dM\,d\hat{\sigma}_
{i+j\to Q\bar{Q}(M)+X}/dM,
\end{equation}
i.e., the open heavy quark cross section is integrated over the 
invariant mass $M$
of the $Q\bar{Q}$ pair up to the open heavy flavour threshold. It is 
then assumed that the quarkonium production cross section is a 
universal (for each $\psi$) fraction $f_\psi$ of the sub-threshold 
cross section. 

Spiritually, the CEM is close to NRQCD in that it also allows a 
$Q\bar{Q}$ pair in a color octet state in the hard collision 
to hadronize into a quarkonium. It is also clear that, because 
the NRQCD expansion arises from an expansion of the open 
$Q\bar{Q}$ production amplitude at small relative $Q\bar{Q}$ 
momentum and because 
$m_{Qq}-m_Q\ll m_Q$, 
the CEM is very similar to NRQCD as far as kinematic 
dependences of the production cross section are concerned. 
The difference arises in the importance that is assigned to the 
various terms that arise in the expansion close to threshold. 
In terms of NRQCD matrix elements this difference can be summarized 
by the statement 
that the power counting implied by the CEM would assign $v^{2 d-6}$ 
to any dimension $d$ operator, independent of the color and spin 
state of the $Q\bar{Q}$ pair (see Tab.~\ref{tab2} and \ref{tab3}). 
The usual argument is that the emission of soft gluons in the 
hadronization of a $Q\bar{Q}$ pair randomizes spin and color, so 
that by the time the quarkonium forms, any information on the 
state of the initial $Q\bar{Q}$ pair has been lost. The problem 
with the argument is that soft gluons do not flip a heavy quark spin 
easily, a piece of information that is incorporated in NRQCD via spin 
symmetry but disregarded in the CEM. Moreover, the CEM treats 
a quarkonium as if its size were that of a light hadron, in which 
case the multipole expansion is not valid. The CEM also does not 
incorporate the heavy quark limit, in which the binding becomes 
Coulombic. This does not impede the color evaporation model to provide 
a description that is phenomenologically successful in some cases, 
in particular as  -- in contrast to the color singlet model -- 
the correct kinematic dependences are incorporated. In general, 
with only one parameter $f_\psi$ the CEM is too restrictive. It 
predicts are universal $\chi/(J/\psi)$-ratio $f_\chi/f_{J/\psi}$, 
which is not supported by the comparison of quarkonium production 
in fixed target collisions, photoproduction and $B$ decay. Note 
that -- as NRQCD -- the CEM is based on a leading-twist 
approximation to the heavy quark production cross section.

\subsection{Quarkonium Polarization}

NRQCD factorization also predicts the polarization of the produced 
quarkonia, although, in general, at the expense of introducing further 
non-perturbative matrix elements that do not appear in the cross 
section summed over all $\psi$ polarization states (`unpolarized cross 
section'). Let me sketch the decomposition of matrix elements 
for the case of a $S$-wave, spin-triplet quarkonium \cite{BEN1,BEN2,bc}. 

After expansion of the $Q\bar{Q}$ production amplitude squared 
in relative momentum, and after decomposing color indices 
into color singlet and a color octet term, and spin indices in a spin 
singlet and triplet term, the cross section can be written as a 
sum (compare with (\ref{factform}))
\begin{equation}
\sum_n C_{n;ij}(\mu) \,\langle {\cal O}^{\psi(\lambda)}_{n;ij} \rangle,
\end{equation}
where the short-distance part $C$ is still coupled to the matrix elements 
through three-vector indices that arise in the expansion in 
relative momentum. To decouple these indices, one writes down the 
most general decomposition of the Cartesian tensor 
$\langle {\cal O}^{\psi(\lambda)}_{n;ij} \rangle$ incorporating 
rotational invariance and the fact that the rest frame 
matrix elements can depend only on the polarization vector (in general, 
tensor) of the quarkonium. 

The constraints from spin symmetry are incorporated as follows. Because 
of spin symmetry, spin and orbital angular momentum are separately 
good quantum numbers at leading order in the velocity expansion. The 
angular momentum  
part of the quarkonium wave function is thus a direct product $S\times  
L_{i}$ ($i$ may be a multi-index for orbital angular momentum greater 
than 1) in 
spin and orbital angular momentum and can be represented as 
\begin{eqnarray}
&& {}^1\!S_0 \qquad 1 \nonumber \\
&& {}^3\!S_1 \qquad \epsilon_a(\lambda)\,\sigma^a\nonumber\\
&& {}^1\!P_1 \qquad \epsilon_i(\lambda)\\
&& {}^3\!P_J \qquad \sum_{\alpha,\rho} \langle J\lambda|1\rho;
1\alpha\rangle\,\epsilon_i(\rho)\epsilon_a(\alpha)\,\sigma^a
\nonumber
\end{eqnarray}
in the quarkonium rest frame, where $\langle J\lambda|1\rho;
1\alpha\rangle$ denotes a Clebsch-Gordon coefficient and 
$\epsilon(\lambda)$ an angular mo\-men\-tum\--one polarization vector. 
A general NRQCD matrix element can then be written as 
\begin{eqnarray}
&&\sum_{X}\langle 0|\chi^\dagger\kappa D_{i_n}\ldots D_{i_n}\psi
|\psi(\lambda)+X\rangle\langle\psi(\lambda)+X|\psi^\dagger 
\kappa' D_{j_n}\ldots D_{j_m}\chi|0\rangle \nonumber\\
&&\hspace*{1cm} = \,\mbox{tr}(\kappa S)\,\mbox{tr}(\kappa' S)\,
L_i L_j\,\xi_{ij i_1\ldots i_n j_1 \ldots j_m},
\end{eqnarray}
where $\kappa$ is a matrix in spin and color indices and 
\begin{eqnarray}
&&\xi_{kl} = a\,\delta_{kl}\nonumber\\
&&\xi_{klmn} = b_0\delta_{kl}\delta_{mn}+b_1\delta_{km}\delta_{ln}+
b_2\delta_{kn}\delta_{lm}
\end{eqnarray}
etc.. $a$, $b_i$ are scalar non-perturbative parameters. Some 
of them can be expressed in terms of those that appear in unpolarized 
production by taking contractions or summing over polarizations. 
For a $S$-wave quarkonium, 
the orbital angular momentum part is trivial and its is 
straightforward to obtain
\begin{eqnarray}
&&\sum_{X}\langle 0|\chi^\dagger \sigma^a\psi
|\psi(\lambda)+X\rangle\langle\psi(\lambda)+X|\psi^\dagger 
\sigma^b\chi|0\rangle = \frac{1}{3}  
\langle {\cal O}_1^{\psi}({}^3\!S_1)\rangle\,\epsilon^{a*}(\lambda)
\epsilon^b(\lambda)\nonumber\\
&&\sum_{X}\langle 0|\chi^\dagger \sigma^a T^A\psi
|\psi(\lambda)+X\rangle\langle\psi(\lambda)+X|\psi^\dagger 
\sigma^b T^a\chi|0\rangle = \frac{1}{3}  
\langle {\cal O}_8^{\psi}({}^3\!S_1)\rangle\,\epsilon^{a*}(\lambda)
\epsilon^b(\lambda)\nonumber\\
&&\sum_{X}\langle 0|\chi^\dagger \sigma^a T^A\left(-\frac{i}{2} 
\stackrel{\leftrightarrow}{D}_i\right)\psi
|\psi(\lambda)+X\rangle\langle\psi(\lambda)+X|\psi^\dagger 
\sigma^b T^A\left(-\frac{i}{2} 
\stackrel{\leftrightarrow}{D}_j\right)\chi|0\rangle \\
&&\hspace*{1cm} =\, 
\langle {\cal O}_8^{\psi}({}^3\!P_0)\rangle\,\delta_{ij}\,
\epsilon^{a*}(\lambda)\epsilon^b(\lambda)\nonumber\\
&&\sum_{X}\langle 0|\chi^\dagger T^A \psi
|\psi(\lambda)+X\rangle\langle\psi(\lambda)+X|\psi^\dagger 
T^A \chi|0\rangle = \frac{1}{3}  
\langle {\cal O}_8^{\psi}({}^1\!S_0)\rangle.\nonumber
\end{eqnarray}
The matrix elements on the right hand side include implicitly 
a sum over polarization as in the definition of Ref.~\cite{BBL}{} and 
in (\ref{me}). 
For the last line zero would be obtained in the spin-symmetry limit. 
The factor $1/3$ here follows from rotational invariance. Note 
that the decomposition of $P$-wave operators in the third and fourth line 
is not diagonal in the angular momentum basis $J J_z$. As a consequence, 
for $S$-wave production, the total angular momentum of an intermediate 
quark pair in a $P$-wave state is not a good quantum number. 
In this basis, interference term between states with different total 
angular momentum have to be included to obtain a factorized form of 
polarized production cross sections \cite{BEN1}. These interference 
terms vanish when all polarizations are summed over. 

Notice that after applying spin symmetry, no new non-perturbative 
matrix elements had to be introduced to describe 
polarized $S$-wave quarkonium production as compared to 
unpolarized production 
at this order in the velocity expansion. For $P$-wave quarkonia, this 
is also true at leading order in $v^2$, but -- in contrast to $S$-wave 
states -- no more for $v^4$ corrections. 
 
\section{Quarkonium production at the Tevatron}

In this present section I discuss the comparison of predictions for 
$S$-wave production with Tevatron data \cite{tevatron,data}, which 
inspired to a large extent the theoretical development described 
above. In Sect.~\ref{fixedtarget} fixed-target data will be revisited 
in the NRQCD approach.

\subsection{Cross sections}
\label{crosstev}

Let me follow the chronology of the development of theory and data. 
When CDF measured, for the first time, separately $J/\psi$ and $\psi'$ 
production not coming from $B$ decay (`prompt'), they found much 
larger cross sections than expected \cite{tevatron}. The expectation then 
was based on the color singlet process
\begin{equation}
\label{singlet}
g+g\to c\bar{c}[{}^3\!S_1^{(1)}]+g:\qquad \alpha_s^3\frac{(4 m_c^2)^2}{p_t^8}
\end{equation}
for the direct production of $S$-wave states, with additional 
contributions to $J/\psi$ from radiative decays of $\chi_{c1}$ and 
$\chi_{c2}$. As indicated, this lowest order process leads to a 
very steep $p_t$-spectrum $d\hat{\sigma}/d p_t^2$ of the short-distance 
cross section. On the other hand at $p_t\gg 2 m_Q$ the quarkonium 
mass can be considered as small and the inclusive production cross 
section, like any single-particle inclusive cross section in QCD, 
should exhibit scaling:  $d\hat{\sigma}/d p_t^2\sim 1/p_t^4$ up 
to logarithms. Such processes can be described by convoluting a 
(properly factorized) parton production cross section with a 
fragmentation function $D_{i\to H}(z)$, where $z$ denotes the 
fraction of $i$'s momentum transferred to the hadron $H$. 
Braaten and Yuan \cite{by} noted that for quarkonia the dependence 
of fragmentation functions on $z$ can be calculated in the NRQCD 
approach. Since the hadronization 
of the $Q\bar{Q}$ pair takes place by emission of gluons with momenta 
of order $m_Q v^2$ in the quarkonium rest frame, 
the energy fraction of the quarkonium relative to 
the fragmenting parton differs from that of the $Q\bar{Q}$ pair 
only by an 
amount $\delta z\sim v^2\ll 1$. They argued that the higher-order 
color singlet process
\begin{equation}
\label{singletfrag}
g+g\to \left[c\bar{c}[{}^3\!S_1^{(1)}]+gg\right] + g:
\qquad \alpha_s^5\frac{1}{p_t^4}
\end{equation}
exceeds (\ref{singlet}) for $p_t>7\,\mbox{GeV}$ as 
relevant to most of the Tevatron data. 
This expectation was born out by detailed analyses \cite{frag}. At this 
point, taking also into account fragmentation production of $P$-wave 
states and their subsequent decay, theory and data seemed to 
agree for $J/\psi$ production, but showed a factor of 30 discrepancy in 
overall normalization for $\psi'$, a discrepancy then referred to as 
$\psi'$-anomaly. The discrepancy in normalization indicated that 
an additional fragmentation contribution had been missed. Braaten 
and Fleming \cite{BF95} suggested that the color octet fragmentation process
\begin{equation}
\label{octetfrag}
g+g\to c\bar{c}[{}^3\!S_1^{(8)}] + g:
\qquad \alpha_s^3\,v^4\,\frac{1}{p_t^4},
\end{equation}
where a gluon fragments as $g\to  c\bar{c}[{}^3\!S_1^{(8)}]$, 
could make up for 
the missing piece. The price to pay is a new parameter 
$\langle {\cal O}_8^{\psi'}({}^3\!S_1)\rangle$ that had to be fitted to 
the data. Because (\ref{octetfrag}) has two powers of $\alpha_s/\pi$ 
less than 
(\ref{singletfrag}), the magnitude of  
$\langle {\cal O}_8^{\psi'}({}^3\!S_1)\rangle$ turned out to 
be consistent with its $v^4$ suppression according 
to the scaling given in Tab.~\ref{tab2}. Subsequently, CDF was able 
to remove $\chi$ feed-down from $J/\psi$ production. The same 
discrepancy as for $\psi'$ appeared in the $J/\psi$ cross section and 
could be explained as for $\psi'$ by color octet fragmentation 
\cite{CAC95}. 

With this explanation of the $\psi'$-anomaly at hand, further studies 
focused on finding additional consistency checks of the 
color octet hypothesis beyond the consistent size of the 
octet matrix element. One such check derives from polarization and 
will be described in the next subsection. Others come from 
different charmonium production processes. Apart from fixed-target 
production discussed in the following section, I do not undertake 
such a comparison in this article, a comparison that would in any case 
be still preliminary. Let me instead continue with the Tevatron 
analysis and discuss the uncertainties in the theoretical prediction. 

\begin{figure}[p]
   \vspace{-2.3cm}
   \epsfysize=14cm
   \epsfxsize=10cm
   \centerline{\epsffile{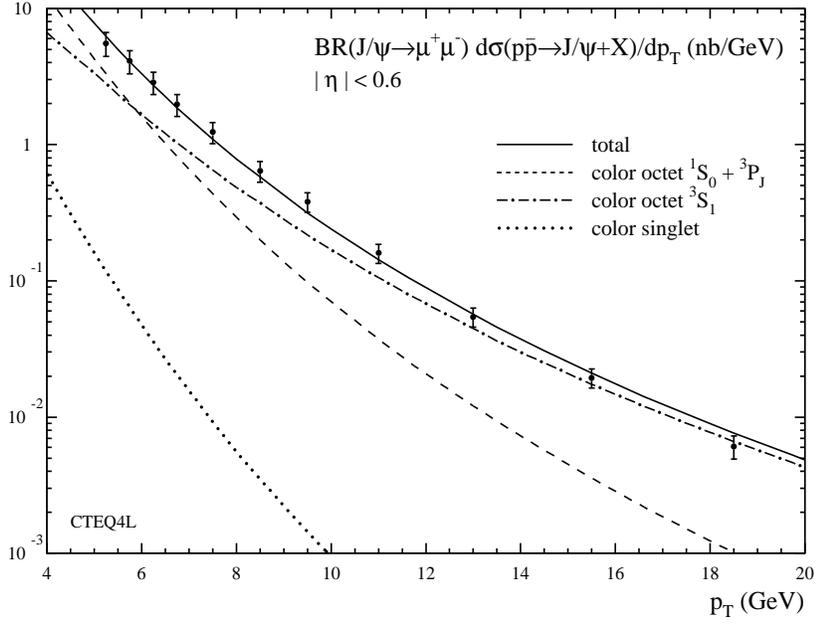}}
   \vspace*{-3cm}
\caption[dummy]{\label{coljpsi} Fit of color octet contributions to direct 
  $J/\psi$ production data from CDF ($\sqrt{s}=1.8\,$TeV,
  pseudo-rapidity cut $|\eta|<0.6$). Theory:
  CTEQ4L parton distribution functions, the
  corresponding $\Lambda_4=235\,$MeV, factorization scale
  $\mu=(p_t^2+4 m_c^2)^{1/2}$ and $m_c=1.5\,$GeV. Taken from 
  Ref.~\cite{bkr}{}.}
\end{figure}

\begin{figure}[p]
   \vspace{-2.3cm}
   \epsfysize=14cm
   \epsfxsize=10cm
   \centerline{\epsffile{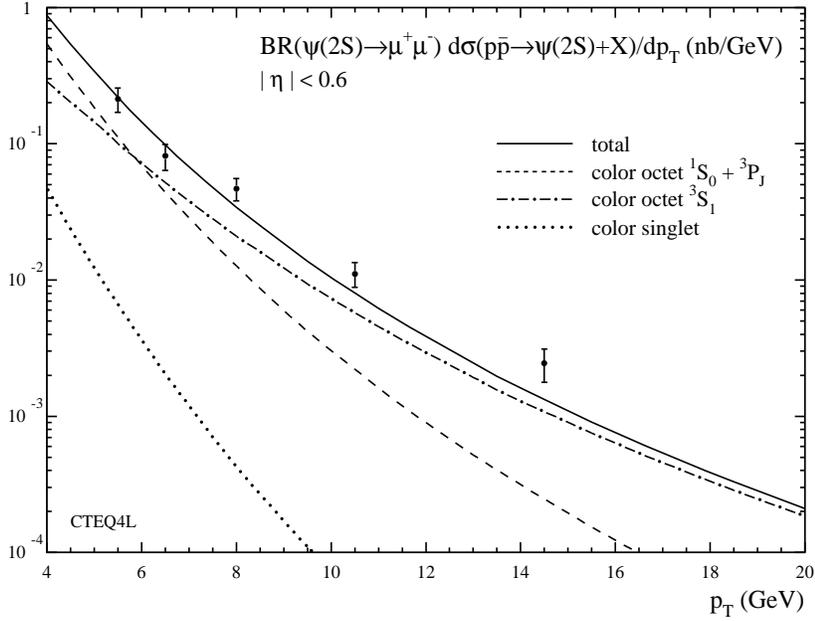}}
   \vspace*{-3cm}
\caption{\label{colpsii} Same as Figure~\ref{coljpsi} 
for prompt $\psi'$ production.}
\end{figure}

At moderate $p_t\sim 2 m_Q$ two further octet production channels, 
which do not have a fragmentation interpretation at order $\alpha_s^3$,
\begin{equation}
\label{octet}
g+g\to c\bar{c}[{}^1\!S_0^{(8)},{}^3\!P_J^{(8)}] + g:
\qquad \alpha_s^3\,v^4\,\frac{4 m_c^2}{p_t^6},
\end{equation}
need to be considered, too, even though they seem to be suppressed 
by $v^4$ with respect to the color singlet process (\ref{singlet}). 
(I'll come back to this point later.) These contributions have been 
calculated in Refs.~\cite{CL,bkr} and turn out to be significant for 
$p_t < 10\,\mbox{GeV}$. In this region (and taking 
$m_c=1.5\,$GeV) the transverse momentum 
distribution is sensitive only to the combination 
$M_{3.5}^\psi(^1\!S_0^{(8)},^3\!P_0^{(8)})$ defined as
\begin{equation}
\label{k}
M_k^\psi(^1\!S_0^{(8)},^3\!P_0^{(8)})\equiv
\langle {\cal O}_8^{\psi}({}^1S_0)\rangle + \frac{k}{m_c^2}\,
\langle {\cal O}_8^{\psi}({}^3P_0)\rangle.
\end{equation}
This matrix element appears as a second fit parameter. The fits 
to the CDF data \cite{data} are shown in Figs.~\ref{coljpsi} and 
\ref{colpsii}. These fits are based on a combined fit of 
$J/\psi$ and $\psi'$ data, keeping the ratios of octet matrix elements 
for $J/\psi$ and $\psi'$ fixed. See Ref.~\cite{bkr} for further details. 
The values of the matrix elements with parameters as specified for 
Fig.~\ref{coljpsi} are found to be
\begin{eqnarray}
\label{numbers}
&&\langle {\cal O}_8^{J/\psi}({}^3S_1)\rangle = 1.06\cdot 10^{-2}\,
\mbox{GeV}^3\nonumber\\
&&\langle {\cal O}_8^{J/\psi}({}^1S_0)\rangle + \frac{3.5}{m_c^2}\,
\langle {\cal O}_8^{J/\psi}({}^3P_0)\rangle = 4.38\cdot 10^{-2}\,
\mbox{GeV}^3\nonumber\\
&&\langle {\cal O}_8^{\psi'}({}^3S_1)\rangle = 0.44\cdot 10^{-2}\,
\mbox{GeV}^3\\
&&\langle {\cal O}_8^{\psi'}({}^1S_0)\rangle + \frac{3.5}{m_c^2}\,
\langle {\cal O}_8^{\psi'}({}^3P_0)\rangle = 1.80\cdot 10^{-2}\,
\mbox{GeV}^3.\nonumber\end{eqnarray}
Note that Figs.~\ref{coljpsi} and \ref{colpsii} include only 
$\alpha_s^3$ processes. The leading color singlet process at large 
$p_t$, Eq.~(\ref{singletfrag}), is not included but would fall 
below the data by a factor of 30 as mentioned above.


It is not surprizing that a good fit of the data can be obtained 
with two contributions to the fit that scale as $1/p_t^4$ and $1/p_t^6$ 
at large $p_t$. It is not at all clear how accurate the fitted 
matrix elements are, however, as any effect that modifies the slope 
of the $p_t$-spectrum would affect the relative weight assigned to the 
two matrix elements $\langle {\cal O}_8^{\psi}({}^3S_1)\rangle$ 
and $M_{3.5}^{\psi}(^1\!S_0^{(8)},^3\!P_0^{(8)})$. Some uncertainties 
that enter the theoretical prediction can be enumerated:\\

1. One may vary the parameters of the calculation such as $\alpha_s$, 
the renormalization and/or factorization scale, as well as the 
parton distribution set. A different value in $\alpha_s$ modifies 
the over-all normalization and also the slope of the $p_t$-distribution, 
because the coupling is evaluated at $\mu=\sqrt{4 m_c^2+p_t^2}$. 

The parametrization of the gluon density affects the prediction 
in a systematic way. Roughly, we may estimate the typical 
proton momentum fractions of the gluons that participate in the hard 
collision as $x_g\sim 2 \sqrt{p_t^2+4 m_c^2}/\sqrt{s}$ for a 
given value of $p_t$. 
A gluon density with steeper small-$x$ behaviour therefore 
steepens the $p_t$-spectrum for the ${}^3\!S_1^{(8)}$ component. 
As a consequence, the $M_{3.5}^{\psi}(^1\!S_0^{(8)},^3\!P_0^{(8)})$ 
component whose magnitude depends crucially on the data being somewhat 
steeper than the ${}^3\!S_1^{(8)}$ component, decreases for a steeper 
gluon density.

The combined effect of a lower $\alpha_s$ and flatter small-$x$ 
gluon distribution is clearly seen in Tab.~\ref{tabme} by 
comparing the value of $M_{3.5}^{\psi}(^1\!S_0^{(8)},^3\!P_0^{(8)})$ 
obtained with MRS(R2) distributions with the one obtained from 
CTEQ4L or GRV. In contrast, the value of 
$\langle {\cal O}_8^{J/\psi}({}^3S_1)\rangle$ is rather insensitive 
to the parton distribution set, but absorbs most of the uncertainty 
in overall normalization obtained by varying the factorization scale 
$\mu$ within a factor two about $\sqrt{4 m_c^2+p_t^2}$ (see the 
second error in Tab.~\ref{tabme}). From these variations alone one 
can conclude that none 
of the two matrix elements is determined to an accuracy better than 
a factor of 2 and that $M_{3.5}^{\psi}(^1\!S_0^{(8)},^3\!P_0^{(8)})$ 
is particularly sensitive to any effect that influence the slope of 
the $p_t$-distribution.\\

\begin{table}[t]
\renewcommand{\arraystretch}{1.25}
$$
\begin{array}{cccc}
\hline\hline
& \mbox{CTEQ4L} & \mbox{GRV (1994) LO} & \mbox{MRS(R2)} \\ 
\hline
\langle {\cal O}_8^{J/\psi}({}^3S_1)\rangle & 
1.06\pm0.14_{-0.59}^{+1.05} & 1.12\pm0.14_{-0.56}^{+0.99} & 
1.40\pm0.22_{-0.79}^{+1.35} \\
M_{3.5}^{J/\psi}(^1\!S_0^{(8)},^3\!P_0^{(8)}) & 
4.38\pm1.15_{-0.74}^{+1.52} & 3.90\pm1.14_{-1.07}^{+1.46} & 
10.9\pm2.07_{-1.26}^{+2.79} \\
\langle {\cal O}_8^{\psi'}({}^3S_1)\rangle & 
0.44\pm0.08_{-0.24}^{+0.43} & 0.46\pm0.08_{-0.23}^{+0.41} & 
0.56\pm0.11_{-0.32}^{+0.54} \\
M_{3.5}^{\psi'}(^1\!S_0^{(8)},^3\!P_0^{(8)}) & 
1.80\pm0.56_{-0.30}^{+0.62} & 1.60\pm0.51_{-0.44}^{+0.60} & 
4.36\pm0.96_{-0.50}^{+1.11}  \\ 
\hline\hline
\end{array}
$$
\caption{\label{tabme}
NRQCD matrix elements in $10^{-2}\,$GeV$^3$. First error statistical, 
second error due to variation of scale. Ratio of $\psi'$ to $J/\psi$ 
fixed. Parton densities from Ref.~\cite{pdfs}{}.}
\end{table}

2. The variation of renormalization and factorization scale gives 
some insight into the magnitude of higher-order corrections in $\alpha_s$ 
only for those contributions that have the same kinematic dependence 
as the leading-order term. In high-energy processes or at not so 
large transverse momentum this may be rather misleading.

Looking at Fig.~\ref{coljpsi} or \ref{colpsii} we may wonder why 
at $p_t\sim 2 m_c$ the contribution from color octet 
${}^3\!P_J$ and ${}^1\!S_0$ $c\bar{c}$ pairs exceeds the color singlet 
contribution by about a factor 20, although from the scaling rules 
the color octet contribution would have been expected to be 
suppressed by $v^4\sim 1/10$. Since the ratio of matrix elements is 
indeed $M_{3.5}^{\psi}(^1\!S_0^{(8)},^3\!P_0^{(8)})/
\langle {\cal O}_1^{\psi}({}^3S_1)\rangle\sim 4\cdot 10^{-2}$, the 
short-distance coefficient of the octet term must be three orders 
of magnitudes larger than the coefficient of the singlet term. 
Examining the short-distance coefficients at $p_t=2 m_c$, 
one finds approximately
\begin{equation}
\frac{d\hat{\sigma}/dp_t^2[{}^3\!P_J^{(8)}]}
{d\hat{\sigma}/dp_t^2[{}^3\!S_1^{(1)}]} \sim 81\,
\frac{\hat{s}^2}{(2 m_c)^4},
\end{equation}
where $\hat{s}$ is the partonic cms energy. The constant 81 is a large 
numerical factor, partly accidental, partly related to color factors and the 
larger number of $P$-wave intermediate states. The parametric ratio 
$\hat{s}^2/(2 m_c)^4$ follows from the fact that diagrams 
with $t$-channel gluon exchange do not contribute to the leading 
order color singlet amplitude. Thus all propagators in the 
color singlet amplitude are off-shell by $\hat{s}$, while the 
octet amplitude is dominated by $t$-channel exchange that allows 
propagators to be off-shell by only $4 m_c^2$. Because the gluon 
density favors small-$x$ gluons, $\hat{s}$ is not tremendously large 
at the Tevatron. Nevertheless, at $p_t=2 m_c$, the lower kinematic 
limit specifies $\hat{s} \geq (1+\sqrt{2})^2 \,(2 m_c)^2\sim 5.8 \,
(2 m_c)^2$ and the ratio $\hat{s}^2/(2 m_c)^4$ provides a 
further significant enhancement of the ampiltude. This rough estimate 
approximately coincides with the factor $10^3$ estimated from 
Fig.~\ref{coljpsi}. 

It follows from this discussion that the next-to-leading 
$\alpha_s^4$ color singlet 
contribution to which $t$-channel exchange diagrams contribute should 
be strongly enhanced as well and lead to large $K$-factors that 
increase with increasing transverse momentum, 
rather similar as in photoproduction \cite{kraemer}. At $p_t\gg 2 m_c$, 
this contribution falls as $1/p_t^6$, slower as the lowest order 
contribution (but still faster than the fragmentation component 
(\ref{singletfrag})) and rather similar in shape to the 
${}^3\!P_J$ and ${}^1\!S_0$ octet components. The value of 
$M_{3.5}^{\psi}(^1\!S_0^{(8)},^3\!P_0^{(8)})$ will therefore decrease 
to compensate for the additional singlet contribution.

Another source of potentially large higher-order corrections is related 
to initial state radiation. These, as does intrinsic transverse 
momentum, would primarily modify the $p_t$-distribution in 
the small-$p_t$ region, up to at least several GeV. Initial state 
radiation could be treated either by analytic resummation 
or estimated by Monte Carlo event generators. A first step in this 
direction has been taken \cite{sanchis}. The bulk effect seems 
to be an enhanced short-distance coefficient that leads to a 
decrease of both fitted color octet matrix elements. For a comparison 
of $\langle {\cal O}_8^{\psi}({}^3S_1)\rangle$ in Ref.~\cite{sanchis} 
and those quoted here and in Ref.~\cite{CL} one has to take into 
account that the evolution of the fragmentation function 
in the ${}^3\!S_1^{(8)}$ channel has not been incorporated 
in Ref.~\cite{sanchis}{}. Evolution depletes the fragmentation function 
at large $z$ and consequently increases the fitted value 
of $\langle {\cal O}_8^{\psi}({}^3S_1)\rangle$ by about a factor 
of 2.\\

3. Apart from higher-order radiative corrections, higher-order 
corrections in $v^2$, the parameter of the non-relativistic expansion,  
can be important close to boundaries of partonic thresholds, 
as discussed in Sect.~\ref{sectfact}. Such a situation occurs 
for the color octet fragmentation function that enters the 
most important fragmentation process (\ref{octetfrag}) at large 
$p_t$. At the scale $2 m_c$ it is given by \cite{BF95}
\begin{equation}
\label{fragfunct}
D_{g\to\psi}(z,2 m_c) = \frac{\pi\alpha_s(2 m_c)}{24 m_c^3}\,
\delta(1-z)\,\langle {\cal O}_8^{\psi}({}^3S_1)\rangle.
\end{equation}
The delta-function clearly neglects that a fraction 
$\delta z\sim v^2$ of the gluon momentum is transferred to light hadrons 
in the `hadronization' of the color octet intermediate state. 
Because of the steep $p_t$-distribution, negligence of this 
non-perturbative softening of the fragmentation function 
introduces a systematic bias \cite{B,MP,L} towards a too low fitted 
value of $\langle {\cal O}_8^{\psi}({}^3S_1)\rangle$. 

The momentum taken by the light degrees of freedom (light hadrons) 
is incorporated into the velocity expansion of NRQCD through 
operators with cms derivatives. Resumming the leading contribution 
results in \cite{BRW}
\begin{equation}
D_{g\to\psi}(z,2 m_c) = \frac{\pi\alpha_s(2 m_c)}{24 m_c^3}\,
\int d y_+\,\delta(1-z-y_+)\,f_\psi[{}^3\!S_1^{(8)}](y_+),
\end{equation}
where
\begin{eqnarray}
f_\psi[{}^3\!S_1^{(8)}](y_+) &=& \\ 
&&\hspace*{-2.4cm}
\sum_{X,\lambda}\langle 0|\chi^\dagger\sigma^i T^a\psi
|\psi(\lambda)+X\rangle
\,\delta\!\left(y_+ - \left[\Lambda\cdot (i D)\right]_+/k_+\right)\,
\langle\psi(\lambda)+X|\psi^\dagger 
\sigma^i T^a\chi|0\rangle.\nonumber
\end{eqnarray}
$D$ denotes a (cms) derivative in the quarkonium rest frame,  
$\Lambda$ the Lorentz boost that transforms it into a moving 
frame, $`+'$ the light-cone `plus'-component of a four-vector and 
$k$ is the four-momentum of the `hadronizing' color octet 
$c\bar{c}$ pair. The distribution $f_\psi[{}^3\!S_1^{(8)}](y_+)$ 
can be interpreted as a distribution 
of light-cone momentum fraction $y_+$ taken by light hadrons 
in the process:\footnote{While writing this review, I learnt 
of related ideas by M.~Mangano, who introduces a similar distribution 
function on phenomenological grounds outside the context of the 
NRQCD expansion.}
\begin{equation}
\label{hadronization}
c\bar{c}[{}^3\!S_1^{(8)}] \to \psi + \,\mbox{light hadrons}.
\end{equation}
After resummation the octet fragmentation function is expressed in 
terms of a new unknown function and all predictivity seems to be 
lost. However, we know that $f_\psi[{}^3\!S_1^{(8)}](y_+)$ should 
have support mainly in a region $y_+\sim v^2$. Moreover, it is a 
universal function that could in principle be extracted from other 
processes not necessarily related to fragmentation. Once such 
a universal function is isolated, one can also try to model it and 
check the consistency of the ansatz in different 
production processes.\\

Summarizing this discussion, I think that regarding the numerical 
values of the matrix elements in (\ref{numbers}) as accurate within 
a factor of 2 or less for $M_{3.5}^{\psi}(^1\!S_0^{(8)},^3\!P_0^{(8)})$ 
would not be overly conservative. It is worth noting that the 
color evaporation model mentioned earlier can reproduce the 
Tevatron data rather reasonably \cite{cemtev}. This is no surprize for 
the shape of the $p_t$-distribution, as 
the CEM includes a fragmentation component. It is less obvious 
that this works (again within factors of 2) with the same values 
of normalization factors $f_\psi$ required to describe fixed-target 
data. From the point of view of NRQCD such agreement would be considered 
coincidental, perhaps related to the fact that the octet matrix elements 
relevant to both processes all scale equally as $v^4$.

Prompt $S$-wave bottomonium production has also been measured by CDF, 
although a $\chi_b$-component could not yet be separated. A comparison 
with predictions can be found in Ref.~\cite{CL}{}. As $v^4\sim 1/100$ 
for bottomonium color octet contributions are less relevant than for 
charmonium in the region $p_t<15\,$GeV of the Tevatron data. Given 
the uncertainties in the color singlet contribution discussed above and 
the octet matrix elements for bottomonium, the necessity of 
color octet contributions is not yet as firmly established as for 
charmonium production. A realistic description of the $p_t$-spectrum 
requires the resummation of soft gluon radiation, which has not yet 
been undertaken.

\subsection{Polarization}
\setcounter{footnote}{0}

Perhaps the most decisive test of the NRQCD picture of quarkonium 
production will come from a polarization measurement of direct 
$J/\psi$ and $\psi'$ at the Tevatron at large quarkonium transverse 
momentum. Recall that at large $p_t$, the production cross section 
is dominated by gluon fragmentation into $c\bar{c}[{}^3\!S_1^{(8)}]$. 
It was first noted by Cho and Wise \cite{CW} that the $c\bar{c}$ 
pair is transversely polarized, because the coupling of longitudinal 
gluons is suppressed by $4 m_c^2/p_t^2$. Furthermore, to leading order 
in the velocity expansion the subsequent transition (\ref{hadronization}) 
takes place via a double electric dipole transition, which does not 
flip the heavy quark spin. Consequently, at large transverse momentum, 
one should observe transversely polarized quarkonia. The polarization 
can be observed in the angular distribution of decay leptons from 
$\psi\to l^+ l^-$, which can be written as 
\begin{equation}
\label{angular}
\frac{d\Gamma}{d\cos\theta}\propto 1+\alpha\,\cos^2\theta,
\end{equation}
with $\theta$ the angle between the
lepton three-momentum in the $\psi$ rest frame and the polarization
axis, the $\psi$ direction in the hadronic cms frame (lab frame). 
Pure transverse polarization would imply $\alpha=1$. 

Corrections to the asymptotic limit come from three sources, 
corresponding to the three small parameters $v^2$, $\alpha_s/\pi$ 
and $4 m_c^2/p_t^2$ that characterize quarkonium production at the 
Tevatron.\footnote{I am ignoring higher-twist corrections which are 
suppressed in the small ratio $\Lambda/p_t$.} 
Since spin symmetry is violated by higher order interaction 
terms in the NRQCD Lagrangian, longitudinal polarization can 
arise if the transition (\ref{hadronization}) proceeds via two 
magnetic dipole transitions. These corrections scale as $v^4$ 
and do not vanish in the limit of large transverse momentum. 
According to the power counting estimate \cite{BEN1}, they could 
reduce $\alpha$ to 0.85.
 
The second source of corrections comes from $\alpha_s$ corrections 
to the fragmentation function, since radiation of a hard gluon can 
change the $c\bar{c}$ pair's polarization. This correction decreases 
with $p_t$ as $1/(\ln p_t)$. All relevant fragmentation 
functions into longitudinally polarized quarkonia have been computed 
\cite{BEN1} and were found to yield about $5\%$ longitudinal 
polarization at $p_t=20\,$GeV. The corrections turned out to be 
rather small, mainly because the fragmentation functions can only 
be softer than the leading term (\ref{fragfunct}) that produces 
transverse charmonia. After convolution with the short-distance 
cross section this leads to suppression of the higher-order 
contribution. If added to the first correction, 
the second still implies that $\alpha>0.65$ at $p_t=20\,$GeV. This 
estimate is based on $\langle {\cal O}_8^{\psi}({}^3\!P_0)\rangle/m_c^2
\approx \langle {\cal O}_8^{\psi}({}^3\!S_1)\rangle$, consistent with 
Tab.~\ref{tabme} (CTEQ4L). 

\begin{figure}[t]
   \vspace{-2.3cm}
   \epsfysize=14cm
   \epsfxsize=10cm
   \centerline{\epsffile{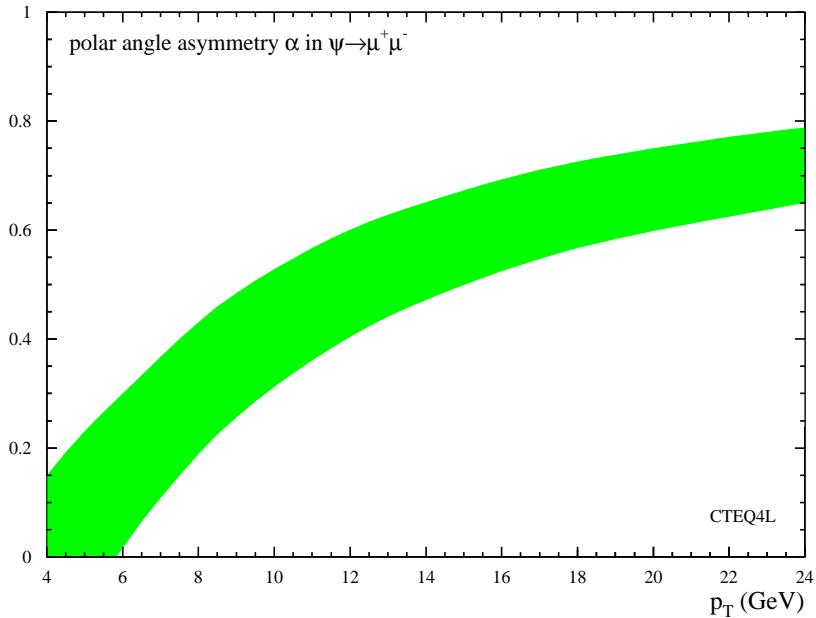}}
   \vspace*{-3cm}
\caption[dummy]{\label{a} $\alpha$ as a function of $p_t$. 
Taken from Ref.~\cite{bkr}{}.}
\end{figure}

Below $p_t\sim 15\,$GeV the dominant source of depolarization 
is due to non-fragmentation contributions \cite{bkr}, because they have 
large short-distance coefficients as seen above. The size of these 
depends crucially on the magnitude of the parameter 
$M_{3.5}^{\psi}(^1\!S_0^{(8)},^3\!P_0^{(8)})$. With the value 
taken from Tab.~\ref{tabme}, the prediction for the parameter 
$\alpha$ in the polar angle distribution is shown in Fig.~\ref{a}. 
The band is based on the non-fragmentation corrections and 
$\alpha_s$ corrections to the fragmentation functions, but 
neglects the depolarization due to spin symmetry breaking of 
order $\delta\alpha\sim 0.1$. It is important that the band 
reflects only the uncertainties in the extraction of NRQCD matrix 
elements due to statistical errors in the data of the unpolarized 
cross section. It does not include any of the theoretical 
uncertainties that could systematically affect the extraction of these 
matrix elements. In particular, if 
$M_{3.5}^{\psi}(^1\!S_0^{(8)},^3\!P_0^{(8)})$ is smaller than 
assumed here, which is not unlikely, the transverse polarization 
fraction would increase. Nevertheless, at $p_t\sim 5\,$GeV, 
Fig.~\ref{a} shows that no remnant of transverse polarization may be 
expected. As $p_t$ increases, the angular distribution becomes 
rapidly more anisotropic. If above $p_t\sim 20\,$GeV 
a substantial fraction of transverse polarization is not observed 
in future (run II) Tevatron data, we will definitely know that 
NRQCD, or at least spin symmetry, is not applicable at the charm mass 
scale. If it is observed, the color evaporation model will have 
lost most of its remaining appeal.
 
\section{Quarkonium production at fixed target}
\label{fixedtarget}

It has long been known, and highlighted most recently in 
Refs.~\cite{SCH94,VAE95}{}, that the color singlet model does not 
give a satisfactory description of all quarkonium production cross 
sections at fixed target energies. The discrepancies arise 
for direct $J/\psi$ and $\chi_1$, both of which can not be 
produced by gluon-gluon fusion at leading order in the color singlet 
channel. When the over-all normalization of the color singlet 
cross section is adjusted to the data, one also predicts a 
significant fraction of transverse polarization, 
$\alpha\approx 0.25$ in the notation of (\ref{angular}) and 
in the Gottfried-Jackson frame. No polarization is observed. 
As compared to the Tevatron `$\psi'$-anomaly' these discrepancies 
are less dramatic and the theoretical uncertainties could 
always be stretched, and higher-twist effects invoked, to make 
them appear even less incisive. Nevertheless, these discrepancies 
must be reassessed in the light of the NRQCD factorization approach.

\subsection{Cross sections}
\label{cross}

Fixed target quarkonium cross sections have been calculated by several 
collaborations \cite{BEN2,TAN95,GUP96}. The spin and color states that can 
be produced in parton-parton collisions at order 
$\alpha_s^2$ are as follows:
\begin{eqnarray}
\label{channels}
gg &\to& c \bar{c}[n]:\quad n={}^1\!S_0^{(1,8)}, {}^3\!P_{0,2}^{(1,8)}, 
{}^1\!P_1^{(8)}, D\mbox{-waves}, \ldots\nonumber\\
q\bar{q} &\to& c \bar{c}[n]:\quad n={}^3\!S_1^{(8)}, D\mbox{-waves}, 
\ldots.
\end{eqnarray}
At order $\alpha_s^3$, after hard gluon radiation, no stringent 
helicity and color constraints remain. Adopting a value 
for $\langle {\cal O}_8^{\psi} ({}^3\!S_1)\rangle$ on the order 
of those shown in Tab.~\ref{tabme}, the quark annihilation channel 
turns out to be insignificant at energies far enough from the 
$c\bar{c}$ threshold such that $x_1 x_2\sim 4 m_c^2/s \ll 1$ 
for the product of parton momentum fractions. The cross sections 
are then dominated by gluon-gluon fusion for both pion and proton 
beams. Keeping the intermediate $c\bar{c}$ states that are leading 
according to the scaling rules of Tab.~\ref{tab2}, the direct 
$J/\psi$ and $\psi'$ production cross section is given by 
\begin{eqnarray}
\label{psicross}
\hat{\sigma}(gg\to\psi') &=& \nonumber\\
&&\hspace*{-3cm}\frac{5\pi^3\alpha_s^2}{12 (2 m_c)^3 s}\,
\delta\!\left(x_1 x_2-\frac{4 m_c^2}{s}\right)
\left[\langle {\cal O}_8^{\psi} ({}^1\!S_0)
\rangle+\frac{3}{m_c^2} \langle {\cal O}_8^{\psi} ({}^3 \!P_0)\rangle
+\frac{4}{5 m_c^2} \langle {\cal O}_8^{\psi} ({}^3 \!P_2)\rangle
\right]\\
&&\hspace*{-3cm} 
+\,\frac{20\pi^2\alpha_s^3}{81 (2 m_c)^5}\,
\Theta\!\left(x_1 x_2-\frac{4 m_c^2}{s}\right)\,\langle 
{\cal O}_1^{\psi} ({}^3 \!S_1)\rangle\,z^2\left[\frac{1-z^2+2 z 
\ln z}{(1-z)^2}+\frac{1-z^2-2z \ln z}{(1+z)^3}\right],\nonumber
\end{eqnarray}
where $z\equiv (2 m_c)^2/(s x_1 x_2)$, $\sqrt{s}$ is the 
center-of-mass energy and $\alpha_s$ is normalized at the scale $2 m_c$. 
The third line gives the color singlet cross section that 
enters at order $\alpha_s^3$. Using spin symmetry to relate 
$\langle {\cal O}_8^{\psi} ({}^3 \!P_J)\rangle = 
(2J+1)\,\langle {\cal O}_8^{\psi} ({}^3 \!P_0)\rangle$, the 
leading color octet contributions are proportional to 
$M_7^\psi(^1\!S_0^{(8)},^3\!P_0^{(8)})$ defined as in (\ref{k}). 

\begin{figure}[t]
   \vspace{0cm}
   \epsfysize=11cm
   \epsfxsize=11cm
   \centerline{\epsffile{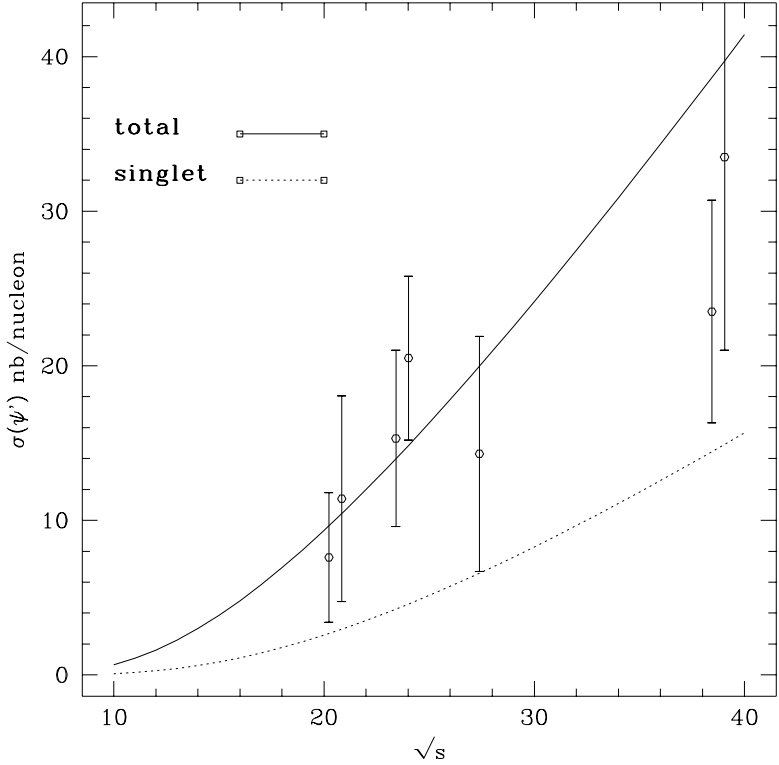}}
   \vspace*{0cm}
\caption{\label{fixpsii} Total (solid) and singlet only (dotted) 
$\psi^\prime$ production cross section ($x_F>0$) in proton-nucleon 
collisions. The solid line is obtained 
with $M_7^{\psi'}(^1\!S_0^{(8)},^3\!P_0^{(8)})=5.2\cdot 10^{-3}\,
$GeV${}^3$.} 
\end{figure} 
\begin{figure}[t]
   \vspace{0cm}
   \epsfysize=11cm
   \epsfxsize=11cm
   \centerline{\epsffile{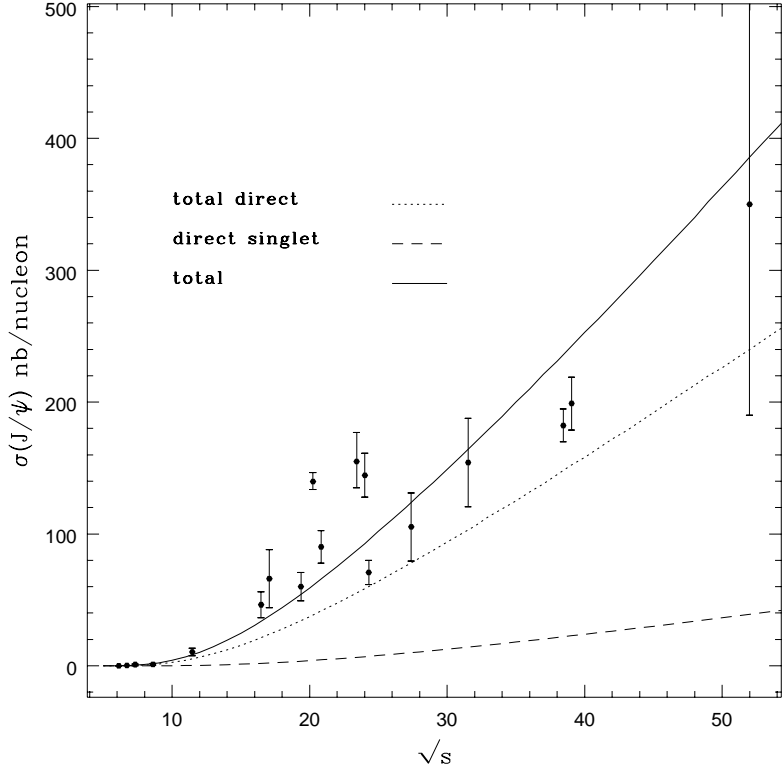}}
   \vspace*{0cm}
\caption{\label{fixjpsi} $J/\psi$ production cross sections ($x_F>0$) in 
proton-nucleon collisions. Solid line: Fit to the total cross section 
including radiative feed-down from the $\chi_{cJ}$ and $\psi'$ with 
$M_7^{J/\psi}(^1\!S_0^{(8)},^3\!P_0^{(8)})=3.0\cdot 10^{-2}\,
$GeV${}^3$. Dashed line: Direct $J/\psi$ cross section.
Dotted line:  Direct $J/\psi$ production, color singlet only.}
\end{figure}

A comparison of the predicted energy dependence and normalization 
of the cross section (including the $q\bar{q}$ annihilation channel) 
with data is shown in Fig.~\ref{fixpsii} and 
\ref{fixjpsi} for $\psi'$ and $J/\psi$ production, respectively. The 
solid curves are obtained with 
\begin{eqnarray}
\label{numbers2}
M_7^{\psi'}(^1\!S_0^{(8)},^3\!P_0^{(8)})\equiv
\langle {\cal O}_8^{\psi'}({}^1S_0)\rangle + \frac{7}{m_c^2}\,
\langle {\cal O}_8^{\psi'}({}^3P_0)\rangle = 0.56\cdot 10^{-2}\,
\mbox{GeV}^3\nonumber\\
M_7^{J/\psi}(^1\!S_0^{(8)},^3\!P_0^{(8)})\equiv
\langle {\cal O}_8^{J/\psi}({}^1S_0)\rangle + \frac{7}{m_c^2}\,
\langle {\cal O}_8^{J/\psi}({}^3P_0)\rangle = 3.0\cdot 10^{-2}\,
\mbox{GeV}^3, 
\end{eqnarray}
with $m_c=1.5\,$GeV, which is assumed in this analysis. The 
color singlet contributions to the direct production cross section 
is shown separately. Note that since the $q\bar{q}$ annihilation channel is 
insignificant, the $P$-wave feed-down incorporated for $J/\psi$ is entirely 
color singlet, because relativistic corrections to 
$\chi_1$ production have been neglected. I will discuss $\chi$ production 
in more detail in Sect.~\ref{chi}. 

Turning first to $\psi'$ production, one notes that given the 
uncertainties inherent to leading order QCD predictions (no $K$-factors 
have been introduced), their renormalization scale-dependence, as well 
as the dependence on $m_c$, the relevance of color octet contributions 
is far from obvious. This is mainly due to the fact that the color 
singlet cross section is dramatically enhanced compared to 
Ref.~\cite{VAE95}, because it is expressed in terms of $2 m_c=3\,$GeV 
rather than the significantly larger $\psi'$ mass. Expressing 
the cross section in terms of quark masses implies a product 
$x_1 x_2$ smaller by $50\%$ than its physical value and enhances the 
cross section as a consequence of the rising gluon density at 
small $x$. As noted earlier, this intrinsic ambiguity of a leading 
order calculation in $v^2$ would be alleviated by resumming 
higher-dimension operators. In the present case it is likely that 
the effect would be a reduction of the color singlet cross section, 
indicating more stringently a missing additional production mechanism. 
The short-distance coefficients of color octet production 
would be likewise affected by implementing the physical phase space 
constraints. In fact, as the invariant mass of the final state 
containing $\psi'$ and light hadrons is even larger than 
$m_{\psi'}$, an even more significant reduction would be expected. 
A fit of the normalization to data would then require a larger 
$M_7^{\psi'}(^1\!S_0^{(8)},^3\!P_0^{(8)})$ than the one 
quoted in (\ref{numbers2}), which would be desirable in 
view of the apparent discrepancy between (\ref{numbers2}) and 
(\ref{numbers}). Given these uncertainties, we can conclude that 
the sizes of color octet contributions as deduced from Tevatron 
and fixed target data are consistent with each other 
as far as order of magnitude is 
concerned and that one could not have hoped for more than that. 

For $J/\psi$, Fig.~\ref{fixjpsi}, we have a clear deficit in the 
color singlet direct production cross section. The difference to 
$\psi'$ comes from the fact that $m_{J/\psi}$ is not much larger 
than $2 m_c$. While this does not remove the ambiguity with the 
choice of $m_c$, one would at least seem to be already closer to the 
physical phase space. If we assume that the $P$-wave feed-down 
is determined by leading order color singlet cross sections, an 
absolute prediction is obtained for the direct $J/\psi$ cross 
section, once the 
unknown parameter $M_7^{J/\psi}(^1\!S_0^{(8)},^3\!P_0^{(8)})$ 
is fitted to the total $J/\psi$ cross section. One finds that 
$\sigma(J/\psi)_{\rm dir}/\sigma(J/\psi)\approx 0.6$ in 
good agreement with experiment. On the other hand this ratio 
comes out a factor 3 too small with color singlet contributions only. 
The result $\sigma(J/\psi)_{\rm dir}/\sigma(J/\psi)\approx 0.6$ 
is still somewhat problematic, because it relies on the assumption 
that the $\chi$ feed-down is accurately reproduced and 
understood. But, as discussed in the following subsection, 
the $\chi_1/\chi_2$ composition of the $\chi$ feed-down assumed 
here conflicts with experiment in that there are too few 
$\chi_1$ predicted. Thus there should be an additional, but neglected 
$\chi$ production mechanism that would then upset the nice 
agreement found for $\sigma(J/\psi)_{\rm dir}/\sigma(J/\psi)$ 
(and make the color singlet model look even worse). This, however, 
can be compensated by taking into account the larger mass 
of $\chi$ states in the same way as for $\psi'$. Thus, it seems 
that color octet contributions in fixed target production are 
natural candidates to remove large discrepancies in over-all 
normalizations, but the remaining uncertainties preclude a 
straightforward test of the universality of color octet matrix elements 
by comparison with Tevatron data. In particular, one can not 
exclude the possibility of significant `higher-twist' contributions 
that would have to be added on top of color octet contributions. 
The numerical situation will be considerably improved by the 
soon available next-to-leading QCD calculation of all relevant 
octet and singlet production channels. For $P$-waves this 
analysis has been reported in Ref.~\cite{MP}{}.

A comparison of cross sections in experiments with pion beams, using 
(\ref{numbers2}) extracted from proton beam experiments shows 
that the former are then systematically under-predicted by a factor 
of about 2. The resolution of this discrepancy, already 
known in the context of color singlet model predictions \cite{SCH94}, 
is clearly outside the scope of NRQCD, which assumes factorization 
of the hadronic initial from the hadronic final state. 

The analysis of bottomonium $S$-wave cross sections suffers from 
the lack of sufficient experimental data, and in particular the 
absence of any experimental information on $\chi_b$ cross sections. 
Using color singlet wave functions and an estimate of color octet 
matrix elements, one finds \cite{BEN2} a larger indirect contribution 
to $S$-wave cross sections than for charmonium. The observation 
of $\Upsilon(3S)$ comparable to $\Upsilon(1S)$ in the E772 experiment 
\cite{ALD91} then indicates the existence of yet undetected 
$\chi_b(3P)$ states below the open bottom threshold.

In addition to integrated cross sections, $x_F$ \cite{GUP962}and $p_t$ 
\cite{slep} 
distributions may be analysed. Since the $x_F$ distribution 
is determined by the gluon flux, the distributions are identical 
for color singlet and octet contributions and do not lead to 
further insight. Note that at large $x_F$ the NRQCD expansion breaks 
down and higher-twist corrections become large \cite{mueller}. A realistic 
comparison of $p_t$ distributions in the range of available 
fixed target data requires yet accounting of soft gluon effects.

\subsection{The $\chi_1/\chi_2$ ratio}
\label{chi}

As noted in (\ref{channels}), the leading-order gluon-gluon fusion 
process can not produce a spin-1 $P$-wave state. Consequently, at energies 
where $q\bar{q}$ annihilation is irrelevant, 
\begin{equation}
\label{r}
R\equiv\frac{\sigma(\chi_1)}{\sigma(\chi_2)} = 
O\!\left(\frac{\alpha_s}{\pi}\right)\sim 0.05\qquad\mbox{(color singlet)},
\end{equation}
with the numerical value from Ref.~\cite{BEN2}{}. The data \cite{chipapers} 
compiled in Tab.~\ref{tabchi} clearly show more copious $\chi_1$ production.

\begin{table}[t]
\addtolength{\arraycolsep}{0.2cm}
\renewcommand{\arraystretch}{1.25}
$$
\begin{array}{c|ccc}
\hline\hline
\mbox{Experiment} & \mbox{beam/target} & \sqrt{s}/\mbox{GeV} & R  \\ 
\hline
\mbox{E673}    & p\mbox{Be} & 19.4/21.7 & 0.24\pm 0.28 \\
\mbox{E705}    & p\mbox{Li} & 23.7      & 0.09    \\
\mbox{E771}    & p\mbox{Si} & 38.8      & 0.34\pm 0.16 \\ 
\hline
\mbox{WA11}    & \pi\mbox{Be} & 18.63  & 0.7\pm 0.2  \\ 
\mbox{E673}    & \pi\mbox{Be} & 18.88  & 0.96\pm 0.64 \\
\mbox{E673}    & \pi\mbox{Be} & 20.55  & 0.9\pm 0.4 \\
\mbox{E705}    & \pi\mbox{Li} & 23.72  & 0.53 \\
\mbox{E672/706} & \pi\mbox{Be} & 31.08 & 0.57\pm 0.19 \\
\hline\hline
\end{array}
$$
\caption{\label{tabchi}
Experimental data on the $\chi_1/\chi_2$ ratio $R$. Data compiled from 
Refs.\cite{chipapers}{}. In view of the experimental errors no attempt has 
been made to rescale older measurements to account for the latest $\chi$ 
branching fractions.}
\end{table}

In the context of the color singlet model, the $\chi_1$ production 
cross section has sometimes been boosted by taking into account an 
exclusive quark-antiquark fusion mechanism \cite{kuehn} 
$q\bar{q}\to \chi_{cJ}$ at order $\alpha_s^4$, see e.g. the cross 
sections displayed in Fig.~16 of Ref.~\cite{SCH94}{}. As this 
mechanism produces $\chi_{cJ}$ states in the ratio 0:4:1, it raises 
$R$ at low energies, especially in $\pi N$ and $\bar{p} N$ collisions. 
The justification for keeping this higher-order 
contribution derives from the presence of two infrared logarithms, 
so that this contribution is of order $\alpha_s^4\ln^2(m_c/\lambda)$, 
where $\ln(m_c/\lambda)$ is interpreted as a non-perturbative parameter. 
This bears some similarity with the infrared logarithm in 
$q\bar{q}\to\chi_{cJ}+g$ which, as discussed, could be interpreted 
as $q\bar{q}\to c\bar{c}[{}^3\!S_1^{(8)}]\to \chi_{cJ}+X$ in the 
framework of NRQCD factorization. This similarity is superficial, though, 
because the double logarithm above is not regularized by the quarkonium 
binding energy and can not be absorbed into a NRQCD matrix element. 
Thus, if it existed, it would contradict NRQCD factorization. However, 
it seems clear that the origin of the double logarithm is the exclusive 
final state considered. The relevant cut diagram at order 
$\alpha_s^4$ contains other cuts related to interfering 
$q\bar{q} g$ final states, which would cancel the infrared logarithms 
in agreement with NRQCD factorization. Thus, there is no 
infrared enhancement in the quark-antiquark fusion mechanism and 
it can not help to raise the $\chi_1/\chi_2$ ratio.

The next candidates are relativistic corrections in the velocity 
expansion \cite{BEN2,GUP3}. Indeed, since they dominate direct 
$J/\psi$ production, that suffers from the same suppression 
as $\chi_1$ in the color singlet gluon-gluon fusion channel, they 
should naturally also be large for $\chi_1$ production. The 
contributions that scale as $v^4$ relative to the leading 
order contribution for $P$-wave production are listed 
in Tab.~\ref{tab3}. (In each channel, there exist $v^2$ corrections 
due to operators with spatial derivatives squared.) To get 
an estimate with no pretence of accuracy, let me consider only 
the octet ${}^3\!P_{J'}$ intermediate states. They yield 
\begin{equation}
\frac{\sigma(\chi_1)\,\mbox{from}\,{}^3\!P^{(8)}_{J'}}
{\sigma(\chi_2)\,\mbox{from}\,{}^3\!P^{(1)}_{2}}
=\frac{15}{8}\,\frac{\langle {\cal O}_8^{\chi_1} ({}^3\!P_2)\rangle 
+15/4 \langle {\cal O}_8^{\chi_1} ({}^3\!P_0)\rangle}
{\langle {\cal O}_1^{\chi_2} ({}^3\!P_2)\rangle}
\sim \frac{1}{3},
\end{equation}
where I assumed that $\langle {\cal O}_8^{\chi_1} ({}^3\!P_2)\rangle/
\langle {\cal O}_1^{\chi_2} ({}^3\!P_2)\rangle\sim 1/10$ according 
to the scaling $v^4\sim 1/10$. If we multiply this ratio by 2 
to account for the intermediate $S$- and $D$-wave states, we obtain 
a $\chi_1$ production cross section an order of magnitude larger 
than (\ref{r}). Since the new production channels contribute also 
to $\chi_2$ production, let me assume that they contribute to 
$\chi_1$ and $\chi_2$ in the ratio 3:5. Then 
\begin{equation}
\label{myr}
R\approx 0.3.
\end{equation}
I emphasize again that this is a very crude estimate.\footnote{The 
short-distance coefficients of all $v^4$ suppressed intermediate 
states have recently been calculated \cite{GUP4}. Nevertheless, 
the significant number of new matrix element makes it difficult 
to improve on the crude estimate given here. The estimate of $R$ given 
in Ref.~\cite{GUP4} is larger than the one quoted here, because 
for unknown reasons the authors choose to omit the dominant 
color singlet channel in $\chi_2$ production. Accounting for it 
makes their estimate consistent with (\ref{myr}).}
The estimate is in agreement 
with the $\chi_1/\chi_2$ ratio measured in $pN$ collisions, 
but is lower than the measurement in 
$\pi N$ collisions. If the trend of larger 
$R$ for pion collisions exhibited by the measurements shown in 
Tab.~\ref{tabchi} is confirmed, the real problem would no longer be the 
$\chi_1/\chi_2$ ratio per se, but the difference between proton and 
pion beam experiments. Unless there is an unknown enhancement 
of the $q\bar{q}$ annihilation channel, such a difference would violate 
the initial/final state factorization of NRQCD and would have to be 
attributed to a `higher-twist' effect. Note that as long as the 
$q\bar{q}$ annihilation channel can be neglected, the $\chi_1/\chi_2$ ratio 
should be approximately energy-independent.

\subsection{Polarization}

Polarization measurements have been performed for both $\psi$  
and $\psi^\prime$ production 
in pion scattering fixed target experiments \cite{AKE93,HEI91}. 
Both experiments observe an essentially flat angular distribution in 
the decay $\psi\to \mu^+ \mu^-$ ($\psi= J/\psi,\psi'$), 
\begin{equation}
\frac{d\sigma}{d\cos\theta }\propto 1+ \alpha \cos^2 \theta\,,
\end{equation}
\noindent where the angle $\theta$ is defined as the angle between 
the three-momentum vector of the positively charged muon and 
the beam axis in the rest frame of the quarkonium. The observed values 
for $\alpha$ are $0.02\pm 0.14$ for $\psi'$, measured at 
$\sqrt{s}=21.8\,$GeV in the region $x_F>0.25$ and 
$0.028\pm 0.004$ for $J/\psi$ measured at $\sqrt{s}=15.3\,$GeV 
in the region $x_F>0$. In both cases the errors are statistical 
only.\footnote{Contemplating Fig.~11 of Ref.~\cite{AKE93} I am 
strongly tempted to assume that the error is misprinted and that 
the measurement should read $0.028\pm 0.04$.}

Compared to these measurements, the color singlet contributions 
alone yield $\alpha\approx 0.25$ for the direct $S$-wave production cross 
section \cite{VAE95}. The polarization yield of octet contributions 
has been considered in Ref.~\cite{BEN2}{}.\footnote{Polarization 
has also been treated in Ref.~\cite{TAN95}{}. However, in the first 
reference of Ref.~\cite{TAN95} and the preprint version of the 
second it was 
assumed that the ${}^1\!S_0^{(8)}$ production channel yields 
pure transverse polarization rather than no polarization, which leads 
to results significantly different from Ref.~\cite{BEN2}{}. This 
has been corrected in the published version of the second reference.} 
The $c\bar{c}[{}^3\!S_1^{(8)}]$ intermediate state yields pure 
transverse polarization, but is insignificant for the total cross 
section, because it occurs only in $q\bar{q}$ annihilation. The 
$c\bar{c}[{}^1\!S_0^{(8)}]$ intermediate state results in 
unpolarized quarkonia while the $c\bar{c}[{}^3\!P_J^{(8)}]$ 
intermediate states prefer transverse to longitudinal polarization 
by a ratio 
6:1. The longitudinal polarization fraction is therefore proportional 
to $M_3^{\psi}(^1\!S_0^{(8)},^3\!P_0^{(8)})$, different from 
the combination that enters the cross section summed over all 
polarizations. Constraining 
$\langle {\cal O}_8^{\psi} ({}^1\!S_0)\rangle$ and 
$\langle {\cal O}_8^{\psi} ({}^3\!P_0)\rangle$ to be positive, 
the range
\begin{equation}
0.15 < \alpha < 0.44
\end{equation} 
is obtained for $\psi'$ production at $\sqrt{s}=21.8\,$GeV. (The 
energy dependence is rather mild.) The lower bound is attained if 
$\langle {\cal O}_8^{\psi'} ({}^3\!P_0)\rangle=0$ and could be as 
low as 0.08, since, as discussed in Sect.~\ref{cross}, the color singlet 
fraction in $\psi'$ production is likely to be lower than that 
shown in Fig.~\ref{fixpsii}. Thus, the prediction can be consistent 
with data within experimental errors, if 
$\langle {\cal O}_8^{\psi'} ({}^3\!P_0)\rangle$ is small 
compared to $\langle {\cal O}_8^{\psi'} ({}^1\!S_0)\rangle$. 
Such a scenario would be realized in case that the multipole 
suppression of $\langle {\cal O}_8^{\psi'} ({}^1\!S_0)\rangle$ 
is not effective ($\lambda\sim 1$ in Tab.~\ref{tab2}).

The discussion for direct $J/\psi$ production follows the $\psi'$ 
case. To compare with the measurement, the polarization yield from 
$\chi$ and $\psi'$ feed-down has to be computed. One then obtains 
\begin{equation}
\label{apsi}
0.31 < \alpha < 0.62,
\end{equation} 
the larger value of $\alpha$ being related to the fact that only 
color singlet mechanisms to $\chi$ production were considered (so that 
the feed-down is almost purely $\chi_2$, in conflict with experiment) 
and that $\chi_2$ is produced only in a helicity $\pm 2$ state 
in gluon-gluon fusion at leading order. The feed-down therefore 
yields an almost purely transversely polarized 
component to the $J/\psi$ cross 
section. Eq.~(\ref{apsi}) is incompatible with the observed flat 
angular distribution. As a prediction it should however be taken with 
due caution. Taking the estimate of Sect.~\ref{chi} literally, 
only one third of the $\chi$ feed-down is due to 
$\chi_2$ produced in a color singlet state, while the 
polarization yield of the remainder is unknown. Thus, it does 
not seem excluded that the total $\chi$ feed-down is largely 
unpolarized, in which case $\alpha$ can be as small as 0.1. 
To resolve this issue, a measurement of $\chi$ polarization would 
be desirable. In particular, one would like to know to what 
extent $\chi_{c2}$ is produced with helicity $\pm 2$. 
Meanwhile, the situation remains unclear and one may speculate 
on the role of higher-twist final state interactions 
without impunity.

The comparison with polarization data, as well as the 
compatibility of matrix elements determined 
from large-$p_t$ and fixed target production, 
Eqs.~(\ref{numbers}) and (\ref{numbers2}) could be ameliorated, if, 
as has been suggested \cite{fleming}, 
$\langle {\cal O}_8^{\psi} ({}^3\!P_0)\rangle$ is actually 
negative. This is indeed possible in principle for the 
renormalized matrix element, because of operator mixing, and 
in particular a quadratic divergence that would appear in 
schemes with an explicit factorization scale. However, 
phenomenologically, a negative value is unacceptable. 
The quadratic divergence implies that in any process 
$\langle {\cal O}_8^{\psi} ({}^3\!P_0)\rangle$ appears as a 
combination 
\begin{equation}
\langle {\cal O}_8^{\psi} ({}^3\!P_0)\rangle + 
k\alpha_s\mu^2 
\langle {\cal O}_1^{\psi} ({}^3\!S_1)\rangle,
\end{equation}
where $\mu$ denotes the factorization scale and the constant 
$k$ follows from separating and including only 
the soft gluon contribution 
to the color singlet cross section. As this combination 
enters a physical cross section, it must be positive in any 
factorization scheme. On the other hand, in all phenomenological  
applications, the higher order color singlet piece is not 
included into the theoretical prediction or is argued to 
be smaller than the octet contribution. Consequently, the 
phenomenologically extracted value of 
$\langle {\cal O}_8^{\psi} ({}^3\!P_0)\rangle$ alone must 
satisfy the positivity constraint. If a fit prefers this 
matrix element to be negative, it must be attributed to 
uncertainties in either theory or data.

\subsection{Beyond NRQCD}

As emphasized in Sect.~\ref{sectfact}, the NRQCD factorization approach 
is leading twist and neglects potential interactions of the intermediate 
heavy quark pair with the beam or target remnants, as well as 
any production mechanism that involves multi-parton interactions in 
the initial state. Quarkonium production cross sections 
in fixed target experiments, where most quarkonia escape at small angle 
with the beam axis, would be especially vulnerable to such 
corrections. In this subsection, I list some of such phenomena, that 
can not be described by NRQCD. For more comprehensive discussions 
and references, see the overviews given by Brodsky \cite{stan} and by  
Hoyer \cite{hoyer}.

{\em Nuclear dependence.} Charmonium production on nuclear targets 
shows suppression with increasing nucleon number, which is parametrized 
by $\sigma=\sigma_0\,A^\alpha$, where $\alpha\approx 0.9-0.92$ for 
cross sections integrated over positive $x_F$. Since $\alpha\approx 1$ 
for open charm production, this nuclear dependence can not be attributed 
to shadowing in parton distributions. NRQCD can not account for such 
a nuclear dependence due to rescattering of a $c\bar{c}$ pair with the 
nucleus. Since for bottomonia $\alpha\sim 0.97$, nuclear suppression is 
consistent with being higher twist. Note that in the nucleus rest 
frame, the charmonium formation time is long in a high energy collision 
and charmonium forms only after the $c\bar{c}$ pair has left the nucleus. 
Since most $J/\psi$ are now believed to originate from a $c\bar{c}$ pair 
originally produced in a color octet state, the rescattering cross 
section is that of an octet rather a singlet $c\bar{c}$ pair. A further 
consequence is that despite their different sizes, the nuclear dependence 
should be the same for $J/\psi$ and $\psi'$ as is indeed observed.

{\em Comover interactions.} A $c\bar{c}$ pair that moves with nearly equal 
velocity as light quarks from the hadron remnants may find it preferable 
to produce open charm rather than charmonium. Comover interactions 
responsible for this effect would be expected to be largest in the 
nuclear fragmentation region ($x_F<0$), because of the larger number 
of potential comovers. Consequently, $\alpha$ should decrease for 
$x_F<0$, an effect that is observed for both charmonium and bottomonium. 

{\em Large-$x_F$ phenomena.} The region $x_F\to 1$ is particularly 
interesting, because the usual hard scattering picture breaks down in 
this region \cite{mueller}. 
One way to visualize this is to note that all partons 
of the projectile have to be correlated in order that all momentum 
can be transferred to a single parton, that subsequently transfers 
all momentum of the projectile to the $c\bar{c}$ pair. Such a correlation 
is rather short-lived and at large $x_F$ its lifetime can 
approach the hard interaction time scale set by $1/m_c$. A hard collision 
at large $x_F$ then takes a snapshot of a compact Fock sate component 
of the projectile, for which multi-parton interactions are not 
suppressed. Such effects may explain the turn-over to longitudinal 
polarization at large $x_F$ seen by the Chicago-Iowa-Princeton 
collaboration, and also a decrease of $\alpha$ towards 0.7 at 
large $x_F$, since the soft parton in the projectile stopped after 
transferring its momentum to the $c\bar{c}$ might then scatter  
only from the surface of the target nucleus. 
If the hadron wave function 
contains an intrinsic charm component (which it certainly does at 
some level), it would also naturally show itself at large $x_F$, 
as the intrinsic $c\bar{c}$ pair preferentially carries most of the 
projectile momentum. Charmonium production due to intrinsic charm 
is suppressed as $\Lambda^2/m_c^2$ (but can be sizeable in 
narrow kinematic regions) and is not included in NRQCD. 

\section{Conclusion}

In this lecture I have tried to convey a topical overview on 
non-relativistic QCD as an effective theory and some of its applications. 
As any effective theory, NRQCD relies on the presence of 
different mass scales, here the quark mass $m_Q$ and the quark's typical 
velocity $m_Q v$ and energy $m_Q v^2$ in a non-relativistic bound 
state. Assuming perturbation theory at the scale $m_Q$, NRQCD allows us 
to organize a calculation 
as an expansion in $\alpha_s(m_Q)$ and $v^2$. NRQCD shares 
many features with heavy quark effective theory and yet has 
a much richer structure owing to the existence of several 
low energy scales $m_Q v$, $m_Q v^2$ and $\Lambda$. As a consequence 
the power counting in the effective theory is also complicated 
and one cannot rely on the dimension of operators alone. The 
velocity scaling rules of NRQCD are simple and unique in the 
Coulombic limit $m_Q v^2\gg \Lambda$. If this limit were realized, 
our use of NRQCD would mimic non-relativistic QED and the bound 
state properties of quarkonia would be amenable to a perturbative 
treatment. As $\Lambda$ is close to any of the two other bound state 
scales for charmonium, the velocity scaling is more delicate (and 
perhaps the most debated issue in NRQCD) and must in principle be 
assessed for every quarkonium state anew.

To describe quarkonium production, NRQCD has to be amalgamated 
with the concept of factorization in perturbative QCD. This 
opens tremendous perspectives on phenomenology, of which only 
a fraction could be reported here. New insights into quarkonium 
production followed from recognizing the importance of color octet 
and fragmentation production mechanisms, both of which are crucial 
in accounting for the charmonium cross sections observed in the 
Fermilab Tevatron experiments.

Nevertheless, NRQCD is not a `Theory of Everything' in 
quarkonium production. The factorization formula for quarkonium 
production can be compared to the factorization theorem for 
Drell-Yan processes, with higher twist corrections due to, for 
example, initial and final state multi-parton interactions  
neglected in both cases. Such corrections may be more or less 
severe dependent on the production process, but would 
affect fixed target and forward photoproduction in particular. 
The comparison of leading twist predictions with fixed target data 
showed significant improvement in all aspects compared to the 
color singlet model, although some features such as $\chi$ production 
and $\psi$ polarization could be made just consistent with data only upon 
straining the theory to its limits, which does not 
appear entirely satisfactory. In any case, even without 
higher twist corrections, theoretical uncertainties remain 
appreciable and complicate a verification of the universality 
of long-distance matrix elements in NRQCD. It seems, at least in 
my opinion, that at this moment there is no discrepancy in 
any process, including those not discussed here, that could 
not be attributed to one or another difficulty in the theoretical 
prediction, and which thus is understood, even if it is not easily
remedied. As concerns fixed target experiments, much clarity could be 
obtained from measurements of $\chi$ polarization, polarization 
of $\Upsilon(nS)$ and separate $\chi_b$ cross sections.

For any theory, the question `Can it be wrong?' is at least as 
important as `Is it correct?'. Since, in the asymptotic limit 
$m_Q\to\infty$, NRQCD is evidently as correct as QCD itself, one 
should ask whether it could generally be inapplicable at the charmonium 
scale. Here, at last, the situation is clear: $\psi'$ or direct 
$J/\psi$ polarization at the Tevatron at large transverse momentum 
will evidence dramatic success or failure of NRQCD for charmonium 
production.\\

{\bf Acknowledgements.} I wish to thank Michael Kr\"amer and Ira 
Rothstein for their collaboration on quarkonia and Gerhard Schuler for 
many discussions and careful reading of the manuscript.


\newpage
\begin{thebibliography}{10}
\bibitem{charmonium}
J.J.~Aubert {\em et al.}, Phys. Rev. Lett. {\bf 33} (1974) 1404;
J.-E.~Augustin {\em et al.}, Phys. Rev. Lett. {\bf 33} (1974) 1406.
\bibitem{BBL} G.T.~Bodwin, E.~Braaten and G.P.~Lepage, 
Phys. Rev. {\bf D51} (1995) 1125.
\bibitem{BF95} E.~Braaten and S.~Fleming, 
Phys. Rev. Lett. {\bf 74} (1995) 3327.
\bibitem{tevatron}
F.~Abe {\em et al.}, Phys. Rev. Lett. {\bf 69} (1992) 3704.
\bibitem{BFY96}
E.~Braaten, S.~Fleming and T.C.~Yuan, Ann. Rev. Nucl. 
Part. Sci. {\bf 46} (1996) 197.
\bibitem{B} M.~Beneke, in `Proceedings of the Second Workshop on 
Continuous Advances in QCD', Minneapolis, M.~Polikarpov (ed.), 
World Scientific, Singapore, 1996, p.~12 [hep-ph/9605462]
\bibitem{QQ} A comprehensive overview as of summer 1996 can be found in 
the Proceedings of the 
Quarkonium Physics Workshop, 
University of Illinois, Chicago, June 1996. 
\bibitem{CL86} W.E.~Caswell and G.P.~Lepage, Phys. Lett. 
{\bf B167} (1986) 437.
\bibitem{KN} T.~Kinoshita and M.~Nio, Phys.~Rev. {\bf D53} (1996) 4909; 
M.~Nio and T.~Kinoshita, CLNS97/1463 [hep-ph/9702218].
\bibitem{koerner} J.G.~K\"orner and G.~Thompson, Phys. Lett. 
{\bf B264} (1991) 185; S.~Balk, J.G.~K\"or\-ner and D.~Pirjol, 
Nucl. Phys. {\bf B428} (1994) 499.
\bibitem{hqet} E.~Eichten and B.~Hill, Phys. Lett. {\bf 234} 
(1990) 511; B.~Grinstein, Nucl. Phys. {\bf B339} (1990) 253; 
H.~Georgi, Phys. Lett. {\bf B240} (1990) 253.
\bibitem{cornell} E.~Eichten {\em et al.}, Phys.~Rev. {\bf D17} 
(1977) 3090.
\bibitem{bc} E.~Braaten and Y.-Q.~Chen, Phys. Rev. {\bf D54} (1996) 3216.
\bibitem{MS} T.~Mannel and G.A.~Schuler, Z. Phys. {\bf C67} (1995) 159.
\bibitem{BRW} M.~Beneke, I.Z.~Rothstein and M.B.~Wise, in 
preparation.
\bibitem{MW} T.~Mannel and S.~Wolf, TTP 97-02 [hep-ph/9701324].
\bibitem{RW} I.Z.~Rothstein and M.B.~Wise, UCSD-97-02 [hep-ph/9702404].
\bibitem{barbieri} R.~Barbieri {\em et al.}, 
Phys. Lett. {\bf B95} (1980) 43; Nucl. Phys. 
{\bf B192} (1981) 61.
\bibitem{bbl2}  G.T.~Bodwin, E.~Braaten and G.P.~Lepage, 
Phys. Rev. {\bf D46} (1992) 1914.
\bibitem{SCH94} G.A.~Schuler, `Quarkonium production and decays', 
CERN-TH-7170-94 [hep-ph/9403387].
\bibitem{csm} C.-H. Chang, Nucl. Phys. {\bf B172} (1980) 425, 
E.L.~Berger and D.~Jones, Phys. Rev. {\bf D23} (1981) 1521,  
R.~Baier and R.~R\"uckl, Phys. Lett. {\bf B102} (1981) 364 and 
Ref.~\cite{SCH94} for a review. 
\bibitem{manohar}
A.V.~Manohar, UCSD/PTH 97-01 [hep-ph/9701294].
\bibitem{ira} B.~Grinstein and I.Z.~Rothstein, UCSD-97-06 
[hep-ph/9703298].
\bibitem{MP} M.~Mangano and A.~Petrelli, CERN-96-293 [hep-ph/9610364], 
to appear in the Proceedings of the Quarkonium Physics Workshop, 
University of Illinois, Chicago, June 1996.
\bibitem{vel}
G.P.~Lepage {\em et al.}, Phys. Rev. {\bf D46} (1992) 4052
\bibitem{labelle}
P.~Labelle, McGill/96-33 [hep-ph/9608491].
\bibitem{err} G.T.~Bodwin, E.~Braaten and G.P.~Lepage, 
erratum to Ref.~\cite{BBL}{}, January 1997.
\bibitem{schuler} G.A.~Schuler, CERN-TH/97-12 [hep-ph/9702230],
to appear in the Proceedings of the Quarkonium Physics Workshop, 
University of Illinois, Chicago, June 1996.
\bibitem{cem} H.~Fritzsch, Phys. Lett. {\bf B67} (1977) 217; 
F.~Halzen, Phys. Lett. {\bf B69} (1977) 105.
\bibitem{BEN1}
M.~Beneke and I.Z.~Rothstein, Phys. Lett. {\bf B372} (1996) 157 
[Erratum: {\em ibid.} {\bf B389} (1996) 789].
\bibitem{BEN2} M.~Beneke and I.Z.~Rothstein, Phys. Rev. 
{\bf D54} (1996) 2005 [Erratum: {\em ibid.} {\bf D54} (1996) 7082];
I.Z. Rothstein, UCSD-96-22 [hep-ph/9609281], to appear in the 
Proceedings of the Quarkonium Physics Workshop, University of Illinois, 
Chicago, June 1996.
\bibitem{data}
M.W.~Bailey (for the CDF coll.), FERMILAB-CONF-96-235-E; 
A.~Sansoni (for the CDF coll.), FERMILAB-CONF-96-221-E;
V.~Papdimitriou (for the CDF coll.), to appear in the 
Proceedings of the Quarkonium Physics Workshop, University of Illinois, 
Chicago, June 1996.
\bibitem{by} E.~Braaten and T.C.~Yuan, Phys. Rev. Lett. 
{\bf 71} (1993) 1671.
\bibitem{frag} M.~Cacciari and M.~Greco, Phys. Rev. Lett. 
{\bf 73} (1994) 1586; 
E.~Braaten, M.~Doncheski, S.~Fleming and M.~Mangano, 
Phys.~Lett. {\bf B333} (1994) 548; D.P.~Roy and K.~Sridhar, 
Phys.~Lett. {\bf B339} (1994) 141.
\bibitem{CAC95}
M.~Cacciari, M.~Greco, M.L.~Mangano and A.~Petrelli,
Phys. Lett. {\bf B356} (1995) 560.
\bibitem{CL}
P.~Cho and A.K.~Leibovich, Phys. Rev. {\bf D53} (1996) 150;
{\bf D53} (1996) 6203.
\bibitem{bkr} M.~Beneke and M.~Kr\"amer, CERN-96-310 [hep-ph/9611218], 
to appear in Phys. Rev. {\bf D}.
\bibitem{pdfs}
M.~Gl\"uck, E.~Reya and A.~Vogt, Z. Phys. {\bf C67} (1995) 433; 
A.D.~Martin, R.G.~Roberts and W.J.~Stirling, Phys. Lett. 
{\bf 387} (1996) 419; H.L.~Lai {\em et al.}, 
Phys. Rev. {\bf D55} (1997) 1280.
\bibitem{kraemer} M.~Kr\"amer, Nucl. Phys. {\bf B459} (1996) 3.
\bibitem{sanchis} B.~Cano-Coloma and M.A.~Sanchis-Lozano, 
IFIC/97-1 [hep-ph/9701210].
\bibitem{L} P.~Ernstr\"om, L.~L\"onnblad and M.~V\"anttinen, 
NORDITA-96-78-P [hep-ph/9612408]. 
\bibitem{cemtev} J.~Amundson {\em et al.}, 
Phys. Lett. {\bf 390} (1997) 323; G.A. Schuler 
and R.~Vogt, Phys. Lett. {\bf B387} (1996) 181.
\bibitem{CW}
P.~Cho and M.B.~Wise, Phys. Lett {\bf B346} (1995) 129.
\bibitem{VAE95} M.~V\"anttinen, P.~Hoyer, S.J.~Brodsky and 
W.-K.~Tang, Phys.~Rev. {\bf D51} (1995) 3332.
\bibitem{TAN95} W.-K.~Tang and M.~V\"anttinen, 
Phys. Rev. {\bf D53} (1996) 6203; Phys. Rev. {\bf D54} 
(1996) 4349.
\bibitem{GUP96} S.~Gupta and K.~Sridhar, Phys. Rev. {\bf D54} 
(1996) 5545.
\bibitem{ALD91} D.M.~Alde {\em et al.}, Phys.~Rev.~Lett. {\bf 66} 
(1991) 2285
\bibitem{GUP962} S.~Gupta and K.~Sridhar, Phys. Rev. {\bf D55} 
(1997) 2650.
\bibitem{slep} L.~Slepchenko and A.~Tkabladze, contribution to the 
3rd German-Russian Workshop on Progress in Heavy Quark Physics, Dubna, 
Russia, May 1996 [hep-ph/9608296].
\bibitem{mueller}  S.J.~Brodsky {\em et al.}, 
Nucl. Phys. {\bf B369} (1992) 519. 
\bibitem{chipapers} Y.~Lemoigne {\em et al.}, Phys.~Lett. {\bf B113} 
(1982) 509; S.R.~Hahn {\em et al.}, Phys. Rev. {\bf D30} 
(1984) 671; D.A.~Bauer {\em et al.}, Phys. Rev. Lett. 
{\bf 54} (1985) 753; L.~Antoniazzi {\em et al.}, Phys. Rev. 
{\bf D49} (1994) 543; V.~Koreshev {\em et al.}, Phys. Rev. Lett. 
{\bf 77} (1996) 4294; K.~Hagan (for the E771 collaboration), 
to appear in the Proceedings of the Quarkonium Physics Workshop, 
University of Illinois, 
Chicago, June 1996.
\bibitem{kuehn} J.H.~K\"uhn, Phys. Lett. {\bf B89} (1980) 385.
\bibitem{GUP3} S.~Gupta and P.~Mathews, TIFR/TH/96-53 
[hep-ph/9609504], to appear in Phys. Rev. {\bf D}.
\bibitem{GUP4}S.~Gupta and P.~Mathews, TIFR/TH/97-08 
[hep-ph/9703370].
\bibitem{AKE93} C.~Akerlof {\em et al.}, Phys.~Rev. {\bf D48} 
(1993) 5067. 
\bibitem{HEI91} J.G.~Heinrich {\em et al.}, Phys.~Rev. {\bf D44}  
(1991) 1909.
\bibitem{fleming} S.~Fleming {\em et al.}, Phys. Rev. {\bf D55} 
(1997) 4098. 
\bibitem{stan} S.J.~Brodsky, SLAC-PUB-7306 [hep-ph/9609415], 
to appear in the Proceedings of the Quarkonium Physics Workshop, 
University of Illinois, 
Chicago, June 1996.
\bibitem{hoyer} P.~Hoyer, NORDITA-97-8-P [hep-ph/9702385], 
talk given at the 2nd ELFE
Workshop, St. Malo, France, September 1996.
\end{thebibliography}
\end{document}